\documentclass[12pt]{article}
\usepackage{amsmath,graphicx}
\usepackage[numbers,sort&compress]{natbib}
\renewcommand{\vec}[1]{\boldsymbol{#1}}
\usepackage{fullpage}
\usepackage{amsfonts}
\usepackage{setspace} 
\usepackage{lineno}
\usepackage{amsmath}
\usepackage{pdflscape}
\usepackage{hyperref}
\usepackage{authblk}
\usepackage[table]{xcolor}
\DeclareMathOperator*{\argmax}{arg\,max}
\title{\sc{When do trajectories matter? Identifiability analysis for stochastic transport phenomena}}

\author[1,2]{Matthew J. Simpson}
\author[3]{Michael J. Plank}
\affil[1]{School of Mathematical Sciences, Queensland University of Technology (QUT), Brisbane, Australia.}
\affil[2]{ARC Centre of Excellence for the Mathematical Analysis of Cellular Systems, QUT, Brisbane,  Australia.}
\affil[3]{School of Mathematics and Statistics, University of Canterbury, Christchurch, New Zealand.}

\date{}

\begin{document}


\maketitle

\begin{abstract}
\setstretch{1.0}  Stochastic models of diffusion are routinely used to study dispersal of populations, including populations of animals, plants, seeds and cells.  Advances in imaging and field measurement technologies mean that data are often collected across a range of scales, including count data collected across a series of fixed sampling regions to characterize population-level dispersal, as well as individual trajectory data to examine at the motion of individuals within a diffusive population.  In this work we consider a lattice-based random walk model and examine the extent to which model parameters can be determined by collecting count data and/or trajectory data.  Our analysis combines agent-based stochastic simulations, mean-field partial differential equation approximations, likelihood-based estimation, identifiability analysis, and model-based prediction.  These combined tools reveal that working with count data alone can sometimes lead to challenges involving structural non-identifiability that can be alleviated by collecting trajectory data.   Furthermore, these tools allow us to explore how different experimental designs impact inferential precision by comparing how different trajectory data collection protocols affects practical identifiability.  Open source implementations of all algorithms used in this work are available on \href{https://github.com/ProfMJSimpson/RandomWalkTrajectories}{GitHub}.
\end{abstract}

\newpage 
\section{Introduction} \label{sec:intro}
Random walk models are widely used to understand cell migration, linking individual motility mechanisms to population-level outcomes~\cite{Berg1983,Codling2008,Plank2025a}. Similar frameworks are used in ecology to study animal movement, foraging and animal-mediated seed dispersal~\cite{Harada1994,Morales2022,Okubo2001}. 

\begin{figure}[htp]
  \centering
\includegraphics[width=1.\textwidth]{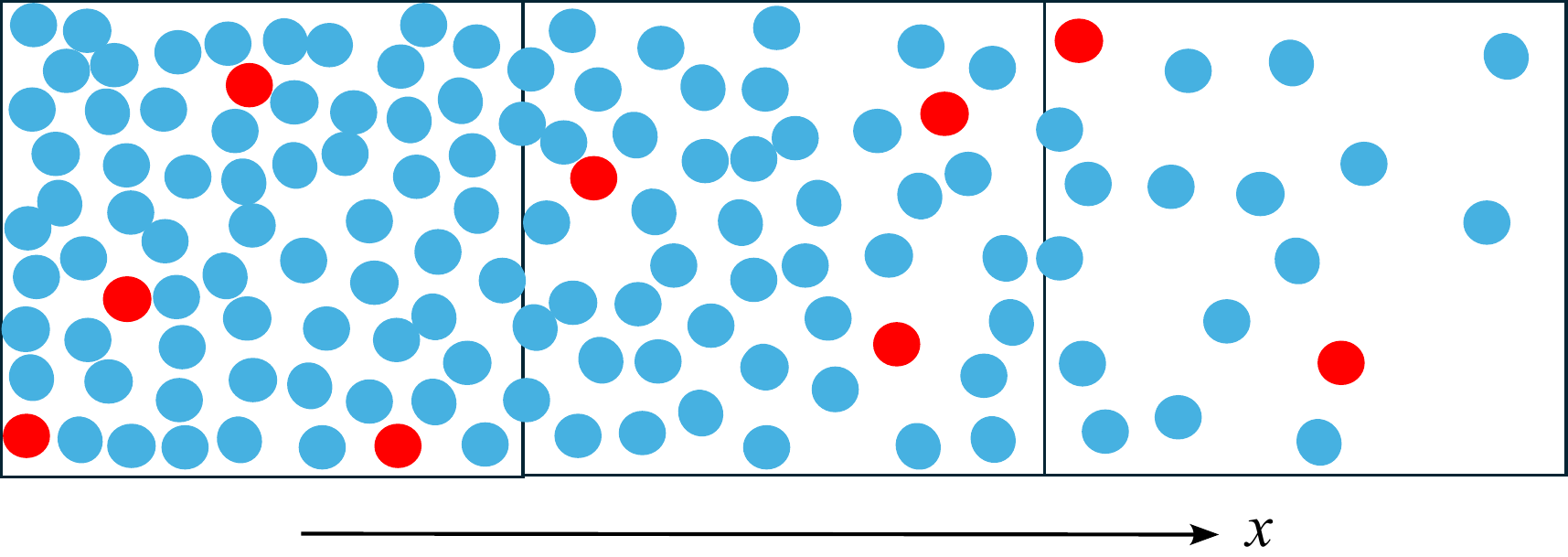}
  \caption{Schematic snapshot of a random walk process involving unlabelled individuals (blue) and a labelled subpopulation (red). Density decreases in the positive $x$-direction, with 68, 47 and 18 individuals counted in the left, central and right regions, respectively. \label{fig:F1}}
\end{figure}

As measurement technologies advance, it is important to understand what different data types reveal about random walk processes. Parameter estimation provides one way to convert empirical measurements into mechanistic understanding. One fundamental measurement is to count individuals in a population undergoing stochastic transport. For example, Figure \ref{fig:F1} shows a schematic population where the number density of individuals clearly decreases with $x$. This trend can be quantified by counting individuals in three non-overlapping rectangular regions and tracking how these counts change over time~\cite{Simpson2013,Treloar2013}. Similar approaches are used in ecology, where field surveys count individuals in grid cells, or \textit{quadrats}~\cite{Seber1986}. 

In addition to count data, there is growing interest in tagging a subset of individuals and tracking their trajectories. This is illustrated in Figure \ref{fig:F1}, where labelled individuals are shown in red. In animal ecology, individuals can be physically tagged~\cite{Perry2015}, while in cell biology fluorescent labels can be used to track individual cells~\cite{Druckenbrod2007}. Despite this interest, relatively little is known about the data quality and quantity required to infer mechanisms from trajectory data.

Trajectory data are common in cell biology, where individually labelled cells are tracked within a bulk population. Such measurements arise in classical two-dimensional culture systems~\cite{Cai2007,Stokes1991a,Young2001}, and in three-dimensional tissues where individual cells are tagged and tracked~\cite{Baker2022,Druckenbrod2005}. In movement ecology, animals are tracked using GPS, telemetry and radar, with random walk models used to interpret dispersal and spatial spreading~\cite{Codling2008,Harris2013,Patterson2008}. Examples include radar-based reconstruction of butterfly flight paths~\cite{Ovaskainen2008} and three-dimensional tracking of bird flocks~\cite{Cavagna2013}. These applications motivate comparing the information content of individual trajectories with population-level observations for estimating transport parameters relevant to diffusion-based model descriptions.

In this work we consider parameter estimation and parameter identifiability for stochastic random walk models, considering both structural and practical identifiability. Parameter identifiability plays a central, but sometimes overlooked role in parameter estimation because it determines whether unknown parameters can be inferred uniquely from measured output. Non-identifiability undermines estimation, interpretation and prediction~\cite{Froehlich2014,Hines2014,Pawitan2001,Simpson2026}.  \textit{Structural identifiability} is a property of the mathematical model under idealised infinite, noise-free data conditions. Structural identifiability addresses the question of whether different parameter values generate different probability distributions of the observable variables. Structural identifiability is typically assessed using Lie derivatives to generate a system of input-output equations, and the solvability properties of this system provide information about structural identifiability~\cite{Chis2011,DiazSeoane2022,Ligon2018}.  \textit{Practical identifiability} is a joint property of the mathematical model and a particular set of data.  Practical identifiability is assessed locally near a given point, such as near the parameter values that provide the best model fit, or more broadly over an extended region parameter space. A common tool for assessing practical identifiability is the profile likelihood that we will use in this work~\cite{Froehlich2014,Raue2009,Simpson2026}.

Standard inference methods for random walk models often rely on likelihood-free, simulation-based methods, such as approximate Bayesian computation. These methods can be computationally demanding~\cite{Browning2017,Lambert2018,Simpson2025}, and can sometimes lead to the neglect of identifiability considerations~\cite{Hines2014,Siekmann2012}. In contrast, likelihood-based parameter estimation is often far more efficient, and provides a platform for profile likelihood-based identifiability analysis~\cite{Froehlich2014,Hines2014,Miles2025,Pawitan2001,Simpson2026}.   Identifiability analysis is often limited to continuum mathematical models where candidate likelihood functions are relatively obvious. In this work we use coarse graining to extend identifiability analysis to stochastic random walk models by using approximate surrogate mean-field descriptions to capture the mean behaviour of the stochastic model.  Stochastic variability is then approximately described using a binomial noise model. 

Very recent work introduced the use of surrogate deterministic models for inference and identifiability analysis using experimentally-motivated count data from stochastic random walk models~\cite{Liu2024,Simpson2025}.  Comparable methods for assessing trajectory data have not been developed. We address this gap using a canonical lattice-based random walk model with biased motion and finite carrying capacity.  Using simulation data, mean-field partial differential equation (PDE) descriptions~\cite{Codling2008,Plank2025a}, and likelihood-based inference~\cite{Pawitan2001,Simpson2025,Simpson2026}, we examine how trajectory data complement count data.  Results indicate that estimating transport parameters from count data for an unbiased motility mechanism can lead to structural non-identifiability~\cite{Hines2014,Simpson2026}. Conversely, estimating transport parameters from trajectory data for biased motility can lead to practical non-identifiability. We also show that these issues can be alleviated by combining count and trajectory data. A key feature of our approach is that estimation, identifiability analysis and prediction are performed using coarse-grained PDE descriptions, making the workflow computationally tractable without repeated stochastic simulations.

A major theme of this work is that diffusive populations are often measured by counting individuals, giving rise to \textit{count} data. A standard approach is to divide these counts by an unknown maximum carrying-capacity count $\kappa$, giving the non-dimensional density $\rho=N/\kappa$, where $\rho=1$ corresponds to maximum packing. Phenomenological continuum models, including exponential growth, logistic growth and ecological competition models, are often formulated in terms of $\rho$ without asking whether $\kappa$ can be estimated from data~\cite{EdelsteinKeshet2005,Kot2001,Murray2002,Skellam1951}. In Figure \ref{fig:F1}, for example, $\kappa$ is not obvious from the count data alone. Here we work directly with counts rather than densities, and ask whether trajectory data can assist in estimating $\kappa$ and other parameters. A surprising outcome is that count data can be insensitive to $\kappa$, whereas trajectory data can estimate $\kappa$ reasonably precisely.

\section{Mathematical Models}

We now outline both the random walk model and briefly derive coarse-grained PDE descriptions of that model.  Stochastic simulation data will be compared with solutions of the PDE  descriptions to provide confidence in the accuracy of the new PDE models.

\subsection{Discrete model}\label{sec:discrete_model}
We consider a discrete time random walk model on a 2D  square lattice with spacing $\Delta$ with time steps of duration $\tau$.  Each site can be occupied by at most $\kappa\in\mathbb{Z}_{>0}$ agents, meaning that the parameter $\kappa$ is a carrying capacity which we treat as a constant, unknown parameter in the model~\cite{Crossley2023,Crossley2024,Taylor2015}.  All simulations are non-dimensional in the sense that we set $\Delta = \tau = 1$, noting that outcomes of these simulations can be re-dimensioned to match any particular choice of length and time scales by re-scaling $\Delta$ and $\tau$ appropriately~\cite{Simpson2025}.  We perform simulations on rectangular domain of height $H$ and width $W = 2L$, so that $ 0 \le y \le H$ and $-L \le x \le L$.  Each site is indexed in the usual way $(i,j)$, and is associated with a coordinate in the Cartesian plane $(x_i,y_j)$ so that  $y_j = (j-1)\Delta$ for $j=1,2,3,\ldots,J$, and $x_i = -L + (i-1)\Delta$ for $i=1,2,3,\dots, I$.

Agent motility is simulated using a random sequential update method~\cite{Chowdhury2005} so that during each time step of duration $\tau$, all agents are given an opportunity to move with probability $M \in [0,1]$.  A motile agent at $(x,y)$ steps to $(x,y \pm \Delta)$ with probability $(1 \pm
\rho_y)/4$, or to $(x \pm \Delta,y)$ with probability $(1 \pm \rho_x)/4$.  Here, $\left| \rho_x \right| \le 1$ and $\left| \rho_y \right| \le 1$ are constant bias parameters that control the degree of motility bias, noting that $\rho_x = \rho_y = 0$ models unbiased motility.  The success of potential motility events depends upon the occupancy status of the target site.  Let $N_{i,j} \in \mathbb{Z}_{\ge0}$ denote the number of agents located at site $(i,j)$.  Potential motility events that would place an agent at site $(i,j)$ are accepted with probability $(1 - N_{i,j}/\kappa)$, and aborted otherwise. This mechanism prevents lattice site from having $N_{i,j} > \kappa$, provided that the initial occupancy of all sites does not exceed $\kappa$.  

The discrete framework that we work with is reasonably flexible.  Other than specifying the initial configurations of agents, we only need to specify four parameters, $\mathbf{\vec{\theta}}_{\textrm{d}} = (M, \kappa, \rho_x,\rho_y)^\top$.  Certain choices of $\kappa$ mean that the discrete model simplifies to some well-known classes of random walk models.  Setting $\kappa=1$ means that the random walk model simplifies to an \textit{exclusion process} which is a kind of random walk that is often implemented for applications where crowding effects are important~\cite{Anguige2009,Bruna2012,Callaghan2006,Mort2016,Painter2002}. In contrast, setting $\kappa$ to be sufficiently large means that sites can be occupied by a many agents, and the random walk model simplifies to a biased Brownian motion in the limit $\kappa \to \infty$~\cite{Codling2008,Plank2025a}.  For intermediate values of $\kappa$ the discrete model interpolates between these two well-known classes of random walk models.

\begin{figure}[htp]
  \centering
\includegraphics[width=1.0\textwidth]{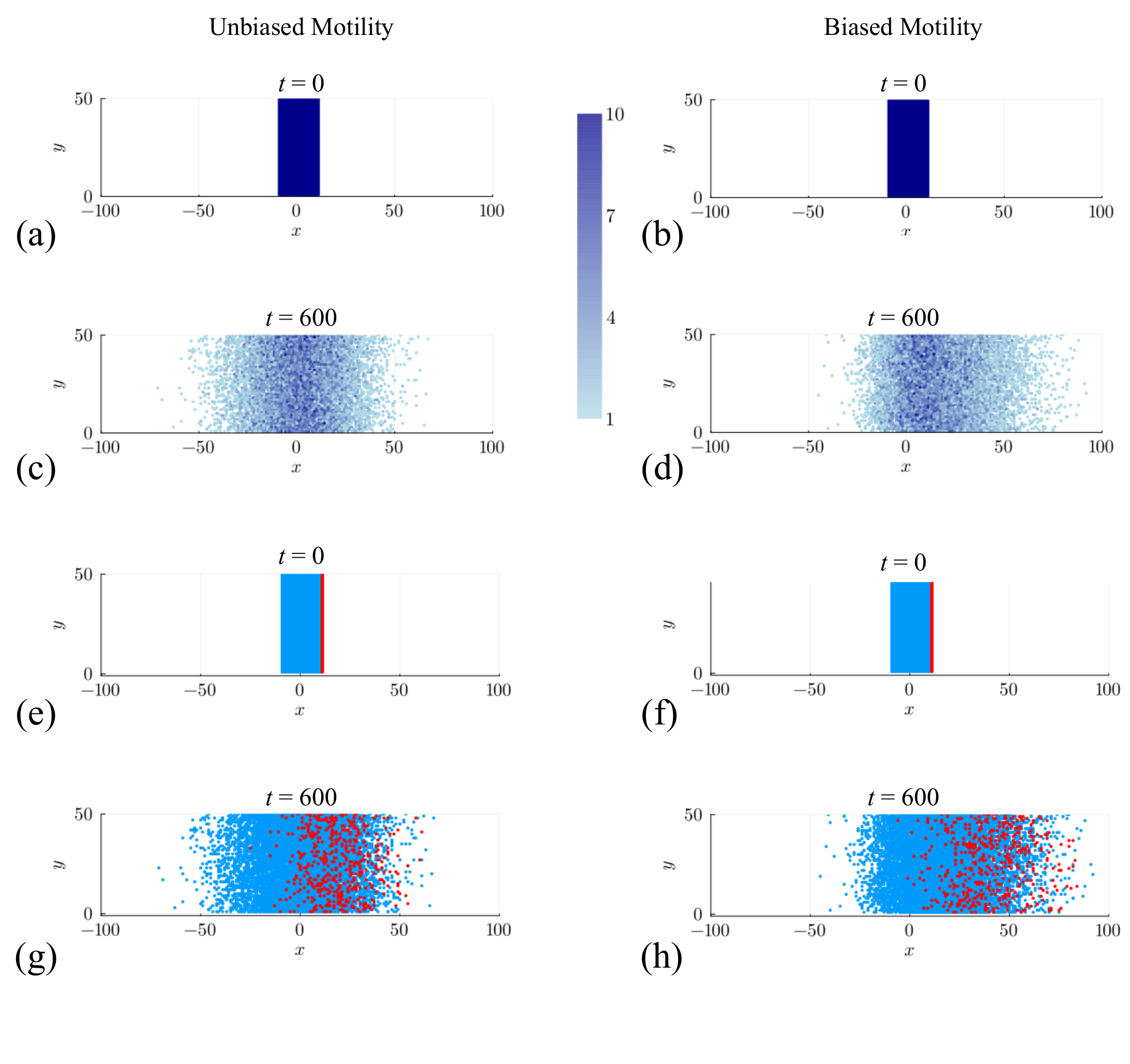}
  \caption{Visualization of the random walk model. The left column shows unbiased motility ($\rho_x=0$), and the right column shows biased motility ($\rho_x=0.1$). (a)--(b) Initial condition with $N_{i,j}(0)=10$ for $|x|\le 10$ and all remaining sites vacant. (c)--(d) Counts per site at $t=600$, with occupancy indicated by the central color bar. (e)--(h) The same simulations showing only occupied sites. Blue discs indicate occupied sites, while red discs indicate sites occupied by tagged agents. Initially, all agents in the right-most occupied column, $x=10$, are tagged. Simulations use $\Delta=\tau=1$, $I=200$, $J=50$, $M=1$ and $\kappa=10$. \label{fig:F2}}
\end{figure}

Typical simulation data are shown in Figure \ref{fig:F2} for unbiased ($\rho_x=\rho_y=0$) and biased ($\rho_x>0$, $\rho_y=0$) motility. Simulations are initialized with 10 agents per site for $|x|\le 10$, with all other sites vacant and all boundaries no-flux; see Figure \ref{fig:F2}(a)--(b). This setup eliminates macroscopic gradients in the vertical direction for $t\ge 0$~\cite{Callaghan2006,Simpson2025}, so net population-level transport occurs in the horizontal direction. This simplification is mathematically convenient and commonly used in cell biology experiments~\cite{Callaghan2006,Simpson2024b}. Figure \ref{fig:F2}(c)--(d) shows the system after $600$ time steps: unbiased motility gives symmetric spreading about $x=0$, whereas biased motility gives an asymmetric distribution.

Figure \ref{fig:F2}(e)--(h) shows the same simulations in terms of occupied sites rather than counts per site. All agents initially in the right-most occupied column, $x=10$, are tagged, corresponding to approximately 5\% of the population. Since each site can contain up to $\kappa=10$ agents, these panels show site occupancy and tagged-agent occupancy rather than individual agents.

\subsection{Surrogate continuum model: Count data}

Rather than relying on repeated stochastic simulations, we approximate the random walk model by a PDE describing its mean behavior~\cite{Codling2008,Plank2025a}. Let $\langle N_{i,j}(t)\rangle\in[0,\kappa]$ denote the average occupancy of site $(i,j)$ at time $t$, estimated over many identically prepared realizations. An approximate conservation statement for the expected change in occupancy over a time step of duration $\tau$ is
\begin{align}\label{eq:NConservation}
\delta \langle N_{i,j}(t) \rangle\ =& \dfrac{M}{4}\left [\langle N_{i-1,j}(t) \rangle \left(1+ \rho_x\right)+\langle N_{i+1,j}(t) \rangle \left(1- \rho_x\right)\right]\left[1 - \dfrac{\langle N_{i,j}(t) \rangle}{\kappa}\right] \\
& + \dfrac{M}{4}\left [\langle N_{i,j-1}(t) \rangle \left(1+ \rho_y\right)+\langle N_{i,j+1}(t) \rangle \left(1- \rho_y\right)\right]\left[1 - \dfrac{\langle N_{i,j}(t) \rangle}{\kappa}\right] \notag \\
& - \dfrac{M}{4}\langle N_{i,j}(t) \rangle \left(1+ \rho_x\right) \left[1 - \dfrac{\langle N_{i+1,j}(t) \rangle}{\kappa}\right]- \dfrac{M}{4}\langle N_{i,j}(t) \rangle \left(1- \rho_x\right) \left[1 - \dfrac{\langle N_{i-1,j}(t) \rangle}{\kappa}\right] \notag \\
& - \dfrac{M}{4}\langle N_{i,j}(t) \rangle \left(1+ \rho_y\right)\left[1 - \dfrac{\langle N_{i,j+1}(t) \rangle}{\kappa}\right]- \dfrac{M}{4}\langle N_{i,j}(t) \rangle \left(1- \rho_y\right) \left[1 - \dfrac{\langle N_{i,j-1}(t) \rangle}{\kappa}\right]. \notag  
\end{align}

Positive terms on the right of  Equation \eqref{eq:NConservation} describe movements onto site $(i,j)$, while negative terms describe movements away from site $(i,j)$. For example, the first term describes movement from site $(i-1,j)$ to site $(i,j)$ under bias in the positive $x$-direction. This term is proportional to the occupancy of the departure site $\langle N_{i-1,j}\rangle/\kappa$, the movement probability $M$, the directional movement probability $(1+\rho_x)/4$, and the probability that the target site accepts the movement, $1-\langle N_{i,j}\rangle/\kappa$. Interpreting the product of these individual probabilities as a net transition probability implies that the occupancy status of sites are independent, which is the classical mean-field approximation. Although questionable for individual events, it is accurate under a wide range of conditions~\cite{Baker2010,Ellner2001}, as we demonstrate later.

To obtain the continuum limit, we divide Equation \eqref{eq:NConservation} by $\tau$, identify $\langle N_{i,j}\rangle$ with $N(x,y,t)$,  and then expand about $(x,y)$ using truncated Taylor series neglecting terms of order $\mathcal{O}(\Delta^3)$. This gives
\begin{equation} \label{eq:N_PDE}
\frac{\partial N}{\partial t} + \nabla \cdot \mathcal{\mathbf{J}}=0, 
\end{equation}
where $\mathbf{J} = \left ( \mathcal{J}_{x}, \mathcal{J}_{y} \right)$ and
\begin{equation}
\mathcal{J}_{x} = -D\dfrac{\partial N}{\partial x}+v_x N\left(1 -  \dfrac{N}{K}\right), \quad 
\mathcal{J}_{y} = -D\dfrac{\partial N}{\partial y}+v_y N\left(1 - \dfrac{N}{K} \right). \notag 
\end{equation}
The carrying capacity $K$, diffusivity $D$, and drift velocity $\mathbf{v}=(v_x,v_y)$ are
\begin{align}
K &= \lim_{\substack{\Delta \to 0 \\ \tau \to 0}} \left( \kappa \right), \quad \quad \quad \quad \quad
D = \lim_{\substack{\Delta \to 0 \\ \tau \to 0}}
\left(\frac{M \Delta^2}{4\tau}\right), \notag \\ 
v_x &= \lim_{\substack{\Delta \to 0 \\ \tau \to 0}}
\left(\frac{M \rho_x \Delta}{2\tau}\right),   
\quad v_y =
\lim_{\substack{\Delta \to 0 \\ \tau \to 0}} \left(\frac{M \rho_y
 \Delta}{2\tau}\right). \notag 
\end{align}
A well-defined continuum limit requires $\Delta^2/\tau=\mathcal{O}(1)$ as $\Delta,\tau\to0$. For biased motility, we also require $\rho_x=\mathcal{O}(\Delta)$ and $\rho_y=\mathcal{O}(\Delta)$ so that $v_x$ and $v_y$ remain $\mathcal{O}(1)$~\cite{Codling2008,Plank2025a}. Thus, the continuum limit is formally valid for sufficiently small bias, although in practice the PDE remains accurate even for maximal bias, $\rho_x=\pm1$ and/or $\rho_y=\pm1$. Finally, some interpretation is required when relating the discrete and continuum carrying capacities since $\kappa$ is a positive integer whereas $K$ is continuous.

The continuum-limit description for $N(x,y,t)$ is given by Equation \eqref{eq:N_PDE}. This model is valid for $K>0$ and reduces to well-known limiting cases. For $K=1$, the fluxes become
\begin{equation*}
\mathcal{J}_{x} = -D\dfrac{\partial N}{\partial x}+v_x N\left(1 -  N\right), \quad
\mathcal{J}_{y} = -D\dfrac{\partial N}{\partial y}+v_y N\left(1 - N \right),
\end{equation*}
corresponding to the continuum limit of an exclusion process, where each lattice site can contain at most one agent~\cite{Plank2025a}. In the limit $K\to\infty$, the fluxes reduce to
\begin{equation*}
\mathcal{J}_{x} = -D\dfrac{\partial N}{\partial x}+v_x N, \quad
\mathcal{J}_{y} = -D\dfrac{\partial N}{\partial y}+v_y N,
\end{equation*}
which is the continuum limit of a biased Brownian random walk with unlimited site occupancy~\cite{Codling2008}.

\subsection{Surrogate continuum model: Trajectory data}

We now derive a macroscopic model for the motion of a tagged agent within the population~\cite{Plank2026}. Suppose one agent is tagged at $t=0$ and tracked through time. Let $P(x,y,t)$ denote the probability density for its location at time $t$, and let $P_{i,j}(t)$ denote the corresponding discrete quantity. As before, $\langle N_{i,j}(t)\rangle$ denotes the expected occupancy of site $(i,j)$. The expected change in $\langle P_{i,j}(t)\rangle$ over a time step of duration $\tau$ is
\begin{align} \label{eq:PConservation}
\delta \langle P_{i,j}(t) \rangle\ =& \dfrac{M}{4}\left [\langle P_{i-1,j}(t) \rangle \left(1+ \rho_x\right)+\langle P_{i+1,j}(t) \rangle \left(1- \rho_x\right)\right]\left[1 - \dfrac{\langle N_{i,j}(t) \rangle}{\kappa}\right] \\
&+ \dfrac{M}{4}\left [\langle P_{i,j-1}(t) \rangle \left(1+ \rho_y\right)+\langle P_{i,j+1}(t) \rangle \left(1- \rho_y\right)\right]\left[1 - \dfrac{\langle N_{i,j}(t) \rangle}{\kappa}\right] \notag \\
&- \dfrac{M}{4}\langle P_{i,j}(t) \rangle \left(1+ \rho_x\right) \left[1 - \dfrac{\langle N_{i+1,j}(t) \rangle}{\kappa}\right]- \dfrac{M}{4} \langle P_{i,j}(t) \rangle \left(1- \rho_x\right) \left[1 - \dfrac{\langle N_{i-1,j}(t) \rangle}{\kappa}\right]\notag \\
&- \dfrac{M}{4} \langle P_{i,j}(t) \rangle \left(1+ \rho_y\right) \left[1 - \dfrac{\langle N_{i,j+1}(t) \rangle}{\kappa}\right]- \dfrac{M}{4}\langle P_{i,j}(t) \rangle \left(1- \rho_y\right) \left[1 - \dfrac{\langle N_{i,j-1}(t) \rangle}{\kappa}\right]. 
\end{align}

As for $N(x,y,t)$, Equation \eqref{eq:PConservation} uses a mean-field approximation assuming independent adjacent site occupancies~\cite{Baker2010,Ellner2001}. Dividing by $\tau$, identifying $\langle P_{i,j}(t)\rangle$ and $\langle N_{i,j}(t)\rangle$ with smooth functions $P(x,y,t)$ and $N(x,y,t)$, expanding neighboring terms in truncated Taylor series, and taking $\Delta,\tau\to0$ with $\Delta^2/\tau=\mathcal{O}(1)$ gives
\begin{equation} \label{eq:P_PDE}
\frac{\partial P}{\partial t} +\nabla \cdot \mathcal{\mathbf{J}}=0,
\end{equation}
where $\mathbf{J} = \left ( \mathcal{J}_{x}, \mathcal{J}_{y} \right)$ and
\begin{align}
\mathcal{J}_{x} &= -D\left(1 -\dfrac{N}{K}\right)\dfrac{\partial P}{\partial x}-D\dfrac{P}{K}\dfrac{\partial N}{\partial x}+ v_xP\left(1 -  \dfrac{N}{K}\right), \notag \\
\mathcal{J}_{y} &= -D\left(1 - \dfrac{N}{K} \right)\dfrac{\partial P}{\partial y} -D\dfrac{P}{K}\dfrac{\partial N}{\partial y}+ v_yP \left(1 - \dfrac{N}{K} \right).  \notag
\end{align}

This PDE describes the probability density of a tagged agent within the bulk population. Thus $P(x,y,t)\in[0,\infty)$ and
\[
\int_{-L}^{L}\int_{0}^{H} P(x,y,t) \, \textrm{d}y \, \textrm{d}x = 1.
\]

The continuum-limit description for $P(x,y,t)$ is given by Equation \eqref{eq:P_PDE} and is valid for $K>0$. As for Equation \eqref{eq:N_PDE}, this model reduces to well-known limiting cases. For $K=1$, the fluxes become
\begin{align}
\mathcal{J}_{x} &= -D\left(1 -N\right)\dfrac{\partial P}{\partial x}-DP\dfrac{\partial N}{\partial x}+ v_xP\left(1 -  N\right), \notag \\
\mathcal{J}_{y} &= -D\left(1 - N \right)\dfrac{\partial P}{\partial y} -DP\dfrac{\partial N}{\partial y}+ v_yP \left(1 - N \right), \notag
\end{align}
which recovers the recently derived result for an exclusion process~\cite{Plank2026}. In contrast, as $K\to\infty$, the fluxes reduce to
\begin{equation*}
\mathcal{J}_{x} = -D\dfrac{\partial P}{\partial x}+ v_xP, \quad 
\mathcal{J}_{y} = -D\dfrac{\partial P}{\partial y}+ v_yP.
\end{equation*}
Thus, in the absence of crowding, the evolution equation for $P(x,y,t)$ has the same form as the evolution equation for $N(x,y,t)$.

\subsection{Simplified surrogate continuum models in one dimension}

As in Figure \ref{fig:F2}, individuals move in any lattice direction, but the initial and boundary conditions eliminate macroscopic gradients in the vertical direction~\cite{Callaghan2006,Simpson2009}. The PDE models for $N(x,y,t)$ and $P(x,y,t)$ therefore reduce to
\begin{align}
\dfrac{\partial N}{\partial t} &= -\dfrac{\partial}{\partial x}\left (-D \dfrac{\partial N}{\partial x} + v N \left[1 - \dfrac{N}{K} \right] \right ), \label{eq:1DN} \\
\dfrac{\partial P}{\partial t} &= -\dfrac{\partial}{\partial x}\left( -D \left[1 - \dfrac{N}{K} \right] \dfrac{\partial P}{\partial x}- D \dfrac{P}{K} \dfrac{\partial N}{\partial x} + v P\left[1 - \dfrac{N}{K} \right] \right), \label{eq:1DP}
\end{align}
for $N(x,t)$ and $P(x,t)$, respectively. For simplicity, we write $v_x$ as $v$. Since these PDEs are nonlinear, we solve them numerically using the method described in the Appendix.

\begin{figure}[htp]
  \centering
\includegraphics[width=1.0\textwidth]{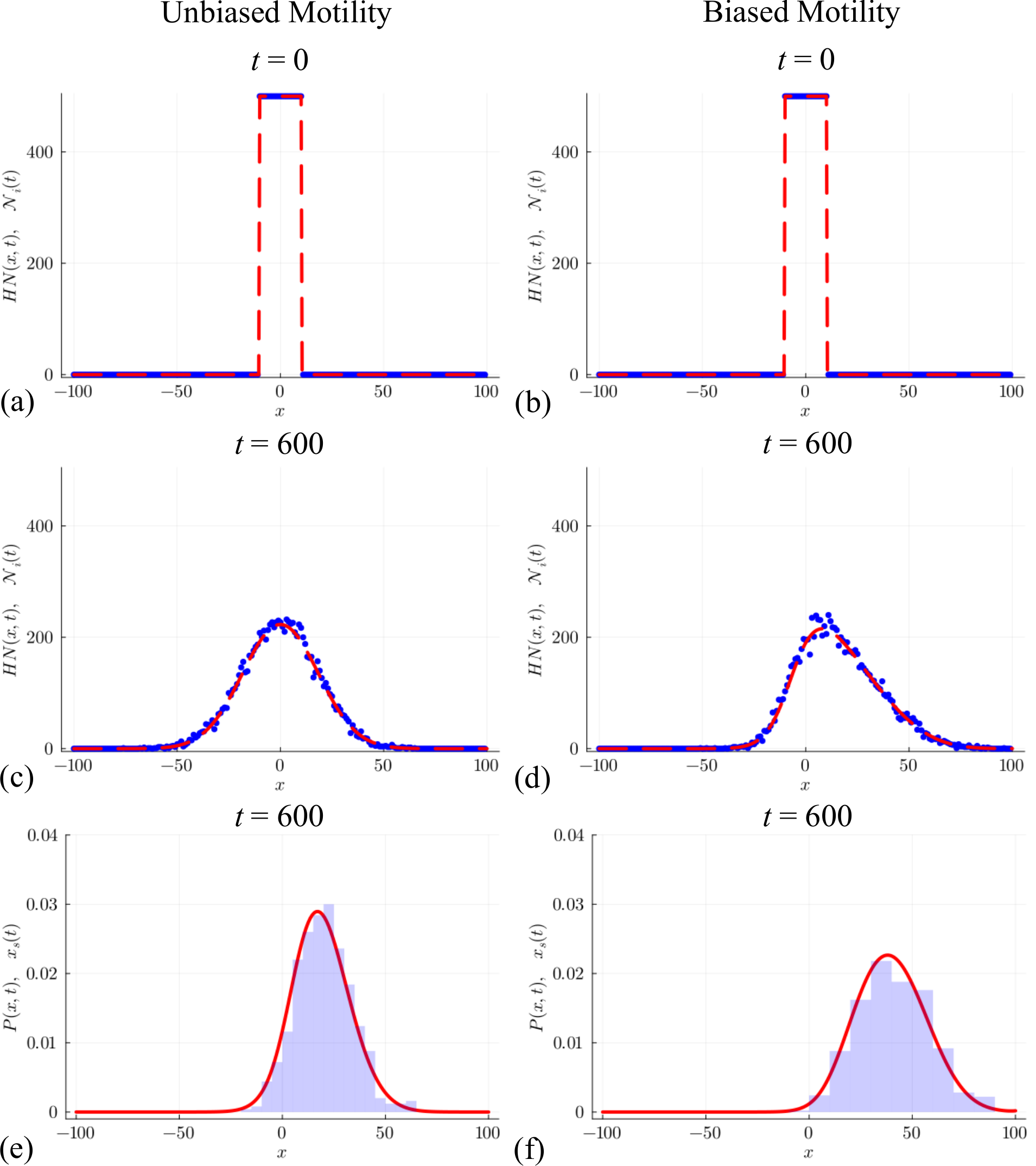}
  \caption{Continuum-discrete comparison for the simulation data in Figure \ref{fig:F2}. (a)--(b) Column counts $\mathcal{N}_i(t)$ (blue dots) and $HN(x,t)$ at $t=0$ (dashed red). (c)--(d) Corresponding profiles at $t=600$. (e)--(f) Normalized histograms of tagged-agent locations from Figure \ref{fig:F2}(g)--(h) (blue) and the PDE solution for $P(x,t)$ at $t=600$ (solid red). \label{fig:F3}}
\end{figure}

We compare data from the discrete model with solutions of Equations \eqref{eq:1DN}--\eqref{eq:1DP}. For count data, we sum the number of agents per site, $n_{i,j}(t)\in[0,\kappa]$, down each column:
\begin{equation}
\mathcal{N}_{i}(t) = \sum_{j=1}^{J} n_{i,j}(t),
\end{equation}
so that $\mathcal{N}_{i}(t) \in \{0,1,2,\ldots, H \kappa\}$. The solution of Equation \eqref{eq:1DN}, $N(x,t)$, describes the average number of agents per site at location $x$. We therefore compare $\mathcal{N}_{i}(t)$ with $H N(x_i,t)$, where $x_i$ is the horizontal location of the $i$th column. Figure \ref{fig:F3}(a)--(d) shows that, for both unbiased and biased simulations from Figure \ref{fig:F2}, Equation \eqref{eq:1DN} predicts the mean trend in the column-based count data but not the stochastic fluctuations.

We then consider tagged agents from Figure \ref{fig:F2}(e)--(h). Let $x_s$, $s=1,\ldots,S$, denote the horizontal locations of the tagged agents. Figure \ref{fig:F3}(e)--(f) shows normalized histograms of $x_s$ at $t=600$. The unbiased and biased simulations use the same number of tagged agents with the same initial locations, but the biased case produces a distribution that is wider and shifted further in the positive $x$-direction. To compare these data with Equation \eqref{eq:1DP}, we set $P(x,0)=0$ except at the initial tagged-agent location $X$, where $P(X,0)=1/\delta$. This gives $\int_{-L}^{L} P(x,0)\,\textrm{d}x=1$, and the no-flux boundary conditions imply $\int_{-L}^{L} P(x,t)\,\textrm{d}x=1$ for $t>0$. Figure \ref{fig:F3}(e)--(f) shows that the PDE solution reasonably approximates the noisy simulation-based histograms.

\section{Inference and Identifiability}

Given the PDE-based surrogates for count and trajectory data, we now explore whether parameters $\vec{\theta}=(D,v,K)^\top$ can be reliably estimated from different data types. We use likelihood-based methods for parameter estimation, identifiability analysis and model-based prediction~\cite{Froehlich2014,Hines2014,Simpson2026}, focusing on count data, trajectory data and their combination.

For count data, suppose column counts $\mathcal{N}_i(t)$ are observed at a single inspection time $t=T$, as in Figure \ref{fig:F3}(c)--(d). Given the solution $N(x,t)$ of Equation \eqref{eq:1DN} with parameters $\vec{\theta}$, define $u(x,t)=N(x,t)/K\in[0,1]$. This is the average occupancy of sites in the column at $x$, relative to the unknown maximum occupancy $K$. We model the $i$th column count as
\begin{equation}
\mathbb{P}(\mathcal{N}_{i}(t) \; | \; \vec{\theta}) = \binom{KH}{\mathcal{N}_{i}(t)}u(x_i,t)^{\mathcal{N}_{i}(t)}\left(1 - u(x_i,t) \right)^{KH - \mathcal{N}_{i}(t)},
\end{equation}
with support $\mathcal{N}_i(t) \in \{0,1,2,\ldots,HK\}$. The corresponding loglikelihood is
\begin{equation}
\ell_{i}^{\textrm{\,c}} (\vec{\theta} \; | \; \mathcal{N}_i(t)) = \log  \binom{KH}{\mathcal{N}_{i}(t)}+ \mathcal{N}_{i}(t) \log \left(u(x_i,t) \right) + \left(KH-\mathcal{N}_{i}(t)\right) \log \left(1 - u(x_i,t) \right),
\end{equation}
where the superscript `c' denotes \textit{count} data. Similar binomial likelihoods have been used for lattice-based random walk models~\cite{Simpson2025} and experimental data~\cite{Simpson2024b}, but with carrying capacity treated as a known, pre-estimated quantity. Here $K$ is estimated, so the binomial coefficient must be retained.

If counts are collected across all columns, and column counts are treated as independent, then
\begin{equation}
\ell^{\textrm{\,c}} (\vec{\theta} \; | \; \mathbf{N}(t)) = \sum_{i=1}^{I} \ell_{i}^{\textrm{\,c}} (\vec{\theta} \; | \; \mathcal{N}_i(t)),
\end{equation}
where $\mathbf{N}(t)=\left(\mathcal{N}_1(t),\mathcal{N}_2(t),\ldots,\mathcal{N}_I(t)\right)^\top$. We evaluate this loglikelihood using the log gamma function~\cite{AbramowitzS1964}, avoiding numerical overflow in factorial terms and allowing non-integer $K$, consistent with treating $K$ as continuous in Equation \eqref{eq:1DN}.

For trajectory data, suppose $S$ agents are tagged at $t=0$, as in Figure \ref{fig:F2}(e)--(h), and their horizontal locations $x_s(t)$ are recorded at $t=T$, for $s=1,\ldots,S$. Given the numerical solution $P(x,t)$ of Equation \eqref{eq:1DP}, the trajectory-based loglikelihood for the $s$th tagged agent is $\ell_{s}^{\textrm{\,t}} (\vec{\theta} \; | \; x_s(t)) = \log  \left (P(x_s,t) \right)$,
where the superscript `t' denotes \textit{trajectory} data. For independent trajectories,
\begin{equation}
\ell^{\textrm{\,t}} (\vec{\theta} \; | \; \mathbf{x}(t)) = \sum_{s=1}^{S} \ell_{s}^{\textrm{\,t}} (\vec{\theta} \; | \; x_s(t)),
\end{equation}
where $\mathbf{x}(t)=\left(x_1(t),x_2(t),\ldots,x_S(t)\right)^\top$.

We use $\ell^{\textrm{\,c}}$ and $\ell^{\textrm{\,t}}$ in the same estimation workflow. For biased motion we estimate $\vec{\theta}=(D,v,K)^\top$, while for unbiased motion, where $\rho_x=v=0$, we estimate $\vec{\theta}=(D,K)^\top$. We will now describe the workflow for $\ell^{\textrm{\,c}}(\vec{\theta}\mid\mathbf{N}(t))$ with three parameters; the same procedure applies to $\ell^{\textrm{\,t}}(\vec{\theta}\mid\mathbf{x}(t))$ for either two or three unknown parameters.

Given data, we evaluate the loglikelihood over a broad region of parameter space containing the true values. For example, for Figure \ref{fig:F2}(d), where $M=1$, $\rho=0.1$ and $\kappa=10$, the true PDE parameters are $\vec{\theta}=(D,v,K)^\top=(0.25,0.05,10)^\top$. We therefore evaluate the loglikelihood over $0.20\le D\le0.30$, $0.04\le v\le0.06$ and $8\le K\le12$ using a uniform $25\times25\times25$ grid. Let $(D_\alpha,v_\beta,K_\gamma)^\top$, for $\alpha,\beta,\gamma=1,\ldots,25$, denote the discretized parameter values. The maximum over this grid approximates the maximum likelihood estimate,
\begin{equation}
\hat{\vec{\theta}}
= \argmax_{\vec{\theta}} \left[ \ell^{\mathrm{\,c}}(\vec{\theta}\mid \mathbf{N}(t)) \right].
\end{equation}
The normalized loglikelihood is
\begin{equation}
\bar{\ell}^{\textrm{\,c}} (\vec{\theta} \; | \; \mathbf{N}(t)) =  \ell^{\textrm{\,c}} (\vec{\theta} \; | \; \mathbf{N}(t)) - \ell^{\textrm{\,c}} (\hat{\vec{\theta}} \; | \; \mathbf{N}(t)),
\end{equation}
so that $\bar{\ell}^{\textrm{\,c}}(\hat{\vec{\theta}}\mid\mathbf{N}(t))=0$.

For two-parameter problems, such as $\vec{\theta}=(D,K)^\top$ under unbiased motility, we plot $\bar{\ell}^{\textrm{\,c}}$ as a heat map to visualize the shape of the loglikelihood near the MLE~\cite{Simpson2026}. We also use Wilks' theorem~\cite{Wilks1938,Royston2007} to define the 95\% confidence threshold $\bar{\ell}^*=-\Delta_{p,q}/2$, where $p$ is the quantile of the $\chi^2$ distribution and $q$ is the number of degrees of freedom. For two unknown parameters, $\bar{\ell}^*=-\Delta_{0.95,2}/2=-2.996$. The contour $\bar{\ell}=\bar{\ell}^*$ bounds the approximate 95\% confidence set. Well-identified parameters have a unique MLE and a constrained confidence set, whereas poorly identified or non-identifiable parameters have a broad confidence set, a poorly defined or non-unique MLE, and/or strong parameter correlations~\cite{Simpson2024}.

For three-parameter problems, such as $\vec{\theta}=(D,v,K)^\top$ under biased motility, we construct bivariate profile loglikelihoods. To achieve this we partition $\vec{\theta}$ into interest parameters $\vec{\psi}$ and nuisance parameters $\vec{\omega}$, so that $\vec{\theta}=(\vec{\psi},\vec{\omega})^\top$~\cite{Hines2014,Simpson2026}. The profile loglikelihood is
\begin{equation}
\bar{\ell}^{\textrm{\,c}}_{\textrm{p}} \left(\vec{\psi} \; | \; \mathbf{N}(t)\right) = \max_{\vec{\omega} \; | \; \vec{\psi}} \; \left[\bar{\ell}^{\textrm{\,c}}(\psi,\omega \; | \;\mathbf{N}(t) )\right],
\end{equation}
which defines the optimal nuisance parameters $\vec{\omega}^*(\vec{\psi})$. For $\vec{\theta}=(D,v,K)^\top$, we construct three bivariate profiles: $\vec{\psi}=(D,v)^\top$ with $\vec{\omega}=K$; $\vec{\psi}=(D,K)^\top$ with $\vec{\omega}=v$; and $\vec{\psi}=(v,K)^\top$ with $\vec{\omega}=D$. Superimposing $\bar{\ell}^*=-2.996$ gives pairwise confidence sets and reveals correlations between parameter pairs. We focus on bivariate profiles because they are easy to visualize and, unlike univariate profiles, directly reveal pairwise parameter correlations.

The profile likelihood functions are straightforward to determine because $\bar{\ell}^{\textrm{\,c}}$ has already been evaluated on a $25^3$ grid. For example, to evaluate the profile for $\vec{\psi}=(D,v)^\top$, we fix each pair $(D_\alpha,v_\beta)$ and maximize over $K_\gamma$, reducing the calculation to a one-dimensional search along a fiber of the three-dimensional array. Although numerical optimization could be used~\cite{Simpson2026}, this grid-based approach is simple, requires no initial estimate, and remains robust for poorly identified problems where iterative solvers can fail~\cite{Simpson2024}. The same procedure gives the trajectory-based profile loglikelihood $\bar{\ell}^{\textrm{\,t}}_{\textrm{p}}(\vec{\psi}\mid\mathbf{x}(t))$.

\section{Results and Discussion}

\subsection{Unbiased motion: When do trajectories matter?}

We begin with the count data in Figure \ref{fig:F3}(c) for unbiased migration and ask whether it identifies $\vec{\theta}=(D,K)^\top$. Since Equation \eqref{eq:1DN} is independent of $K$, the mean-field solution $N(x,t)$ cannot identify $K$, indicating structural non-identifiability~\cite{Hines2014,Simpson2026}. In our likelihood-based approach, however, the PDE solution is combined with the binomial noise model, so $\ell^{\textrm{\,c}}(\vec{\theta}\mid\mathbf{N}(t))$ depends weakly on $K$ through the noise model. Evaluating $\ell^{\textrm{\,c}}$ on a uniform $25\times25$ discretization of $(D,K)$ gives the approximate MLE $\hat{\vec{\theta}}=(0.250,9.167)^\top$, close to the true value. 

\begin{figure}[htp]
  \centering
\includegraphics[width=1\textwidth]{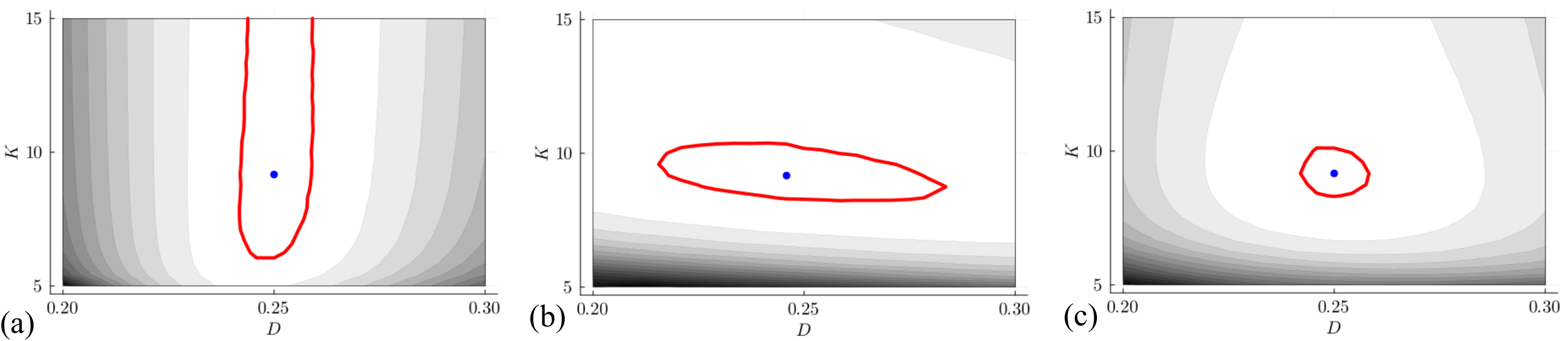}
  \caption{Estimation and identifiability for unbiased motility using data from Figure \ref{fig:F2}. (a)--(c) Heat maps of $\bar{\ell}^{\textrm{\,c}}(\vec{\theta}\mid\mathbf{N}(t))$, $\bar{\ell}^{\textrm{\,t}}(\vec{\theta}\mid\mathbf{x}(t))$ and $\bar{\ell}^{\textrm{\,c+t}}(\vec{\theta}\mid\mathbf{N}(t),\mathbf{x}(t))$, respectively. Each heat map shows the 95\% threshold contour $\bar{\ell}^*=-2.996$ in red. MLEs are $(0.250,9.167)^\top$, $(0.246,9.167)^\top$ and $(0.250,9.167)^\top$ (blue dots), respectively. The grayscale darkens as $\bar{\ell}$ decreases, with shading scaled separately in each panel. \label{fig:F4}}
\end{figure}

This point estimate alone does not quantify uncertainty, so Figure \ref{fig:F4}(a) shows $\bar{\ell}^{\textrm{\,c}}(\vec{\theta}\mid\mathbf{N}(t))$ with the 95\% threshold contour. The confidence set indicates that $D$ is well identified by count data, whereas $K$ is not: the loglikelihood is relatively flat in the $K$ direction. Thus, although the MLE is close to the true value, a broad range of $K$ values match the data almost equally well. This is consistent with Equation \eqref{eq:1DN} being independent of $K$, with only weak dependence introduced through the binomial likelihood.

We next estimate $\vec{\theta}=(D,K)^\top$ using the trajectory data in Figure \ref{fig:F2}(g). Unlike Equation \eqref{eq:1DN}, Equation \eqref{eq:1DP} depends explicitly on both $D$ and $K$. Evaluating $\ell^{\textrm{\,t}}(\vec{\theta}\mid\mathbf{x}(t))$ on the same $(D,K)$ grid gives $\hat{\vec{\theta}}=(0.246,9.167)^\top$, again close to the true values. Figure \ref{fig:F4}(b) shows that the MLE is much better defined than for count data: the 95\% confidence set is contained within the parameter region considered. Thus, for this example, trajectory data identify both $D$ and $K$ reasonably well, whereas count data identify only $D$.

Finally, we combine count and trajectory data. Assuming conditional independence given the model parameters, the combined likelihood is the product of the count and trajectory likelihoods, giving
\begin{equation}
\ell^{\textrm{\,c+t}} (\vec{\theta} \; | \; \mathbf{N}(t),\mathbf{x}(t)) = \ell^{\textrm{\,c}}(\vec{\theta} \; | \; \mathbf{N}(t)) +\ell^{\textrm{\,t}}(\vec{\theta} \; | \; \mathbf{x}(t)), \label{eq:Combined Loglikelihood}
\end{equation}
where the superscript `c+t' denotes combined \textit{count} and \textit{trajectory} data. Using $\ell^{\textrm{\,c+t}}$, the approximate MLE is $\hat{\vec{\theta}}=(0.250,9.167)^\top$. Figure \ref{fig:F4}(c) shows that combining both data types further restricts the 95\% confidence set relative to using either data type alone. Thus, while count data alone provides little information about $K$, adding trajectory data gives greater certainty in the parameter estimates.

\subsection{Biased motion: When do trajectories matter?}

\begin{figure}[htp]
  \centering
\includegraphics[width=1.0\textwidth]{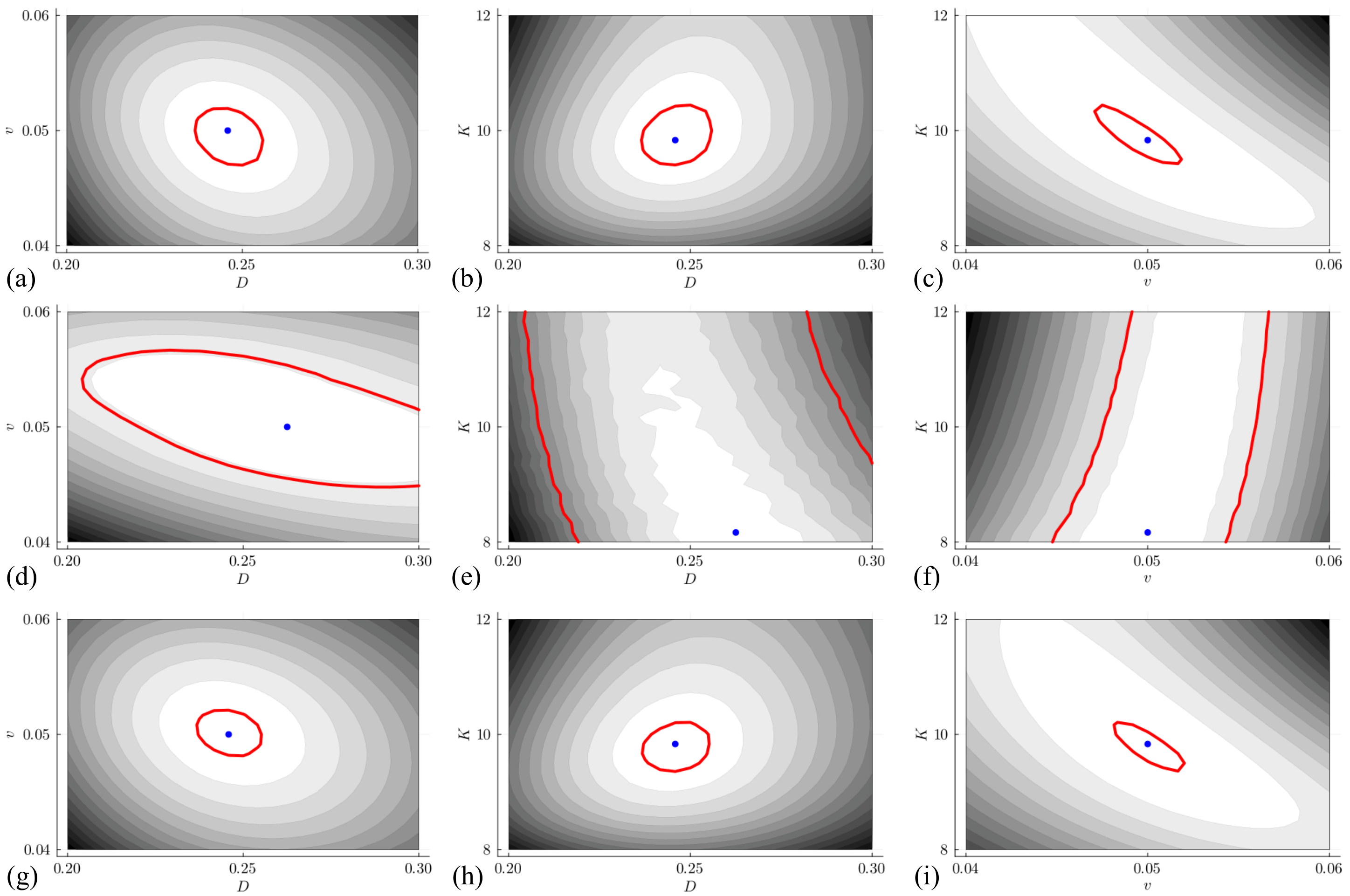}
  \caption{Estimation and identifiability for biased motility using data from Figure \ref{fig:F2}. Heat maps show bivariate profile loglikelihoods using count data in (a)--(c), trajectory data in (d)--(f), and combined count and trajectory data in (g)--(i). Each heat map shows the 95\% threshold contour $\bar{\ell}^*=-2.996$ in red. MLEs (blue dots) are $\hat{\vec{\theta}}=(0.246,0.050,9.833)^\top$ for count data, $\hat{\vec{\theta}}=(0.263,0.050,8.167)^\top$ for trajectory data, and $\hat{\vec{\theta}}=(0.246,0.050,9.833)^\top$ for combined data. The grayscale darkens as $\bar{\ell}$ decreases, with shading scaled separately in each panel. \label{fig:F5}}
\end{figure}

We now consider the count data in Figure \ref{fig:F3}(d) for biased migration and ask whether it identifies $\vec{\theta}=(D,v,K)^\top$. Unlike unbiased migration, when $v\ne0$ Equation \eqref{eq:1DN} depends on all three parameters. Evaluating $\ell^{\textrm{\,c}}(\vec{\theta}\mid\mathbf{N}(t))$ on a uniform discretization of $(D,v,K)$ gives the approximate MLE $\hat{\vec{\theta}}=(0.246,0.050,9.833)^\top$, close to the true values. The bivariate profile likelihoods in Figure \ref{fig:F5}(a)--(c) show that all three parameters are well identified by count data, with the 95\% confidence set tightly constrained around the MLE. The profiles indicate that $D$ is relatively uncorrelated with $v$ and $K$, while $v$ and $K$ show a modest negative correlation, as seen in Figure \ref{fig:F5}(c).

We repeat the analysis using the trajectory data in Figure \ref{fig:F2}(h), evaluating $\ell^{\textrm{\,t}}(\vec{\theta}\mid\mathbf{x}(t))$ on the same $(D,v,K)$ discretization. This gives $\hat{\vec{\theta}}=(0.263,0.050,8.167)^\top$, which is further from the true values than the count-based MLE. The bivariate profiles in Figure \ref{fig:F5}(d)--(f) show that the 95\% confidence regions are also much larger than those for count data. Unlike the unbiased case in Figure \ref{fig:F4}, count data now give more accurate and precise estimates than trajectory data. Combining both data types using $\ell^{\textrm{\,c+t}}(\vec{\theta}\mid\mathbf{N}(t),\mathbf{x}(t))$ gives $\hat{\vec{\theta}}=(0.246,0.050,9.833)^\top$ and tighter confidence sets, as shown in Figure \ref{fig:F5}(g)--(i).

\subsection{Experimental design: How do trajectories matter?}

\begin{figure}[htp]
  \centering
\includegraphics[width=1.0\textwidth]{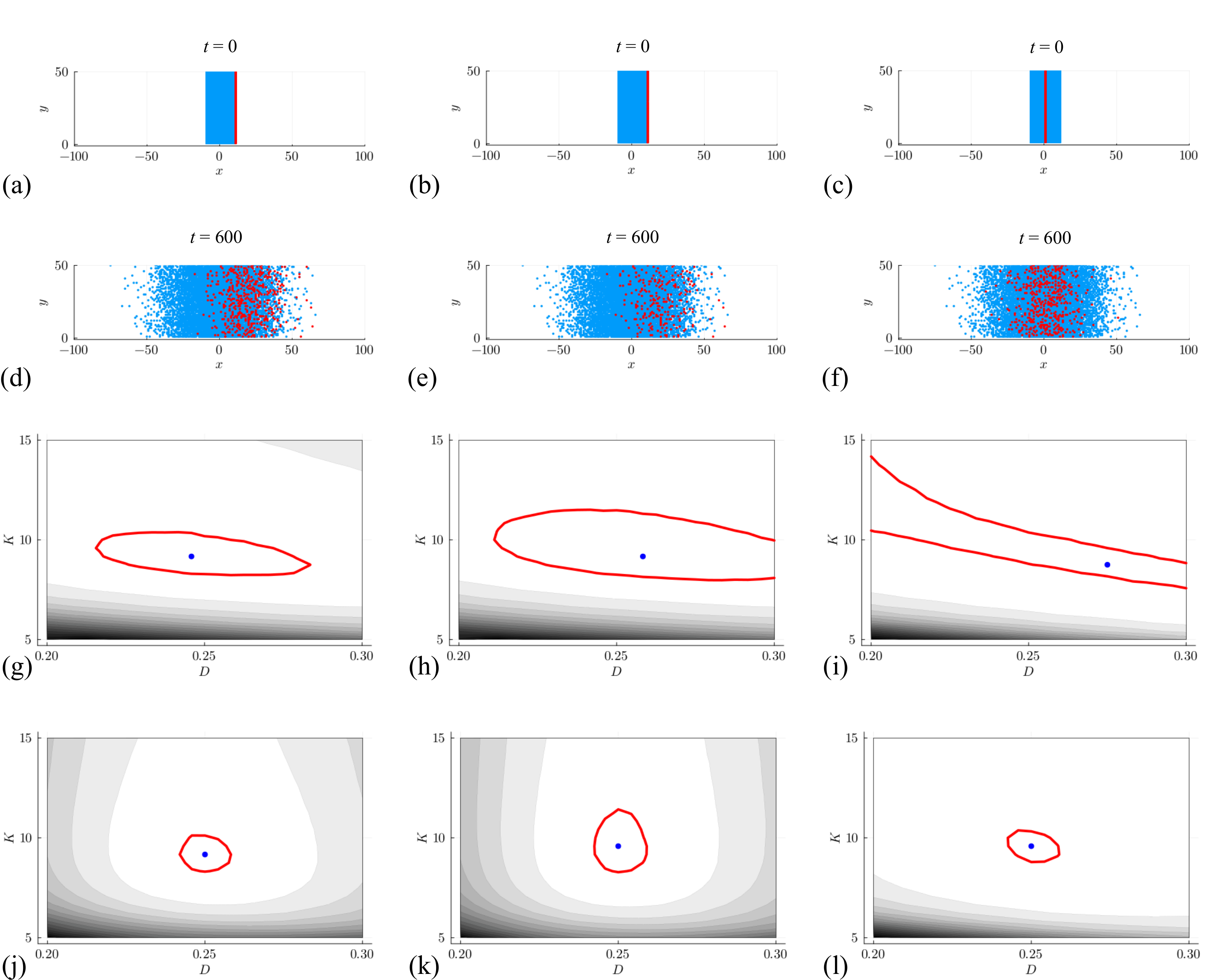}
  \caption{Experimental design with different numbers and configurations of tagged agents for the unbiased random walk in Figure \ref{fig:F2}. The left column tags all agents at $x=10$, the central column tags 50\% of agents at $x=10$, and the right column tags all agents at $x=0$. (a)--(c) and (d)--(f) show snapshots at $t=0$ and $t=600$, respectively. (g)--(i) show heat maps of $\bar{\ell}^{\textrm{\,t}}(\vec{\theta}\mid\mathbf{x}(t))$, with MLEs $\hat{\vec{\theta}}=(0.246,9.167)^\top$, $(0.258,9.167)^\top$ and $(0.275,8.750)^\top$, respectively. (j)--(l) show heat maps of $\bar{\ell}^{\textrm{\,c+t}}(\vec{\theta}\mid\mathbf{N}(t),\mathbf{x}(t))$, with MLEs $\hat{\vec{\theta}}=(0.250,9.167)^\top$, $(0.250,9.583)^\top$ and $(0.250,9.583)^\top$, respectively. Each heat map shows the 95\% threshold contour $\bar{\ell}^*=-2.996$ in red. The grayscale darkens as $\bar{\ell}$ decreases, with shading scaled separately in each panel. \label{fig:F6}}
\end{figure}

Results in Figures \ref{fig:F4}--\ref{fig:F5} use the initial configuration in Figure \ref{fig:F2}(e)--(f), where the population occupies all sites with $|x|\le 10$ and all agents in the right-most occupied column, $x=10$, are tagged. This single configuration allowed us to study parameter identifiability for a fixed tagging protocol. We now vary the number and initial position of tagged agents, while leaving the underlying stochastic transport unchanged. For simplicity, we consider unbiased motility with $\rho_x=v=0$. The left-most column of Figure \ref{fig:F6} repeats the previous design: all agents at $x=10$ are tagged. Snapshots in Figure \ref{fig:F6}(a) and (d) show the initial and final configurations, while Figure \ref{fig:F6}(g) and (j) show $\bar{\ell}^{\textrm{\,t}}(\vec{\theta}\mid\mathbf{x}(t))$ and $\bar{\ell}^{\textrm{\,c+t}}(\vec{\theta}\mid\mathbf{N}(t),\mathbf{x}(t))$, respectively. These reproduce Figure \ref{fig:F4}(b)--(c) and provide a baseline for comparison.

The central column of Figure \ref{fig:F6} uses the same stochastic simulation, except that only half of the agents at $x=10$ are tagged. Comparing Figure \ref{fig:F6}(d) and (e) shows fewer tagged agents. The trajectory loglikelihood in Figure \ref{fig:F6}(h) is less peaked at the MLE than in Figure \ref{fig:F6}(g), and the 95\% confidence region is correspondingly larger. The combined-data loglikelihood in Figure \ref{fig:F6}(k) is also slightly less peaked than in Figure \ref{fig:F6}(j).

We next consider a different tagging location, motivated by cell biology experiments in which cells are tagged at different positions within a population. For example, Druckenbrod and Epstein~\cite{Druckenbrod2007} compared trajectories of cells originating at the leading edge with trajectories of cells originating well behind it. We therefore repeat the same simulation but tag all agents in the central column, $x=0$, as shown in Figure \ref{fig:F6}(c) and (f). Thus, the only difference between Figure \ref{fig:F6}(d) and (f) is that the former tags agents at the leading edge, $x=10$, whereas the latter tags agents within the bulk population, $x=0$. The trajectory loglikelihood in Figure \ref{fig:F6}(i) shows poor identifiability: the 95\% confidence set extends beyond the parameter region considered, indicating that many parameter combinations match the trajectory data. Comparing Figure \ref{fig:F6}(g) and (i) confirms that trajectories originating at the leading edge provide more precise parameter estimates than trajectories originating within the bulk. Combining count and trajectory data, Figure \ref{fig:F6}(l), again constrains the 95\% confidence set to a small region around the MLE.

Poorly defined MLEs, such as in Figure \ref{fig:F6}(i), motivate our grid-based likelihood calculation and fiber-search profiling. Here the loglikelihood is relatively flat across a broad region of parameter space, so iterative optimization can struggle to converge and may be sensitive to the initial estimate of $\vec{\theta}$. In contrast, discretizing the loglikelihood and searching along fibers avoids these difficulties and is attractive for modest problems involving 2--4 parameters~\cite{Simpson2024}.

\subsection{Likelihood-based prediction}

Given a normalized loglikelihood function and parameter confidence set, we can quantify how parameter uncertainty translates into prediction uncertainty by constructing prediction intervals~\cite{Simpson2025,Vardeman1992}. To demonstrate this approach, we focus on the unbiased motility results in Figure \ref{fig:F2}(e) and (g), where all individuals in the column at $x=10$ are tagged. As discussed in relation to Figure \ref{fig:F4}, count data alone identifies $D$ but not $K$. In contrast, trajectory data alone, Figure \ref{fig:F4}(b), and combined count and trajectory data, Figure \ref{fig:F4}(c), identify both $D$ and $K$. Although likelihood-based prediction is usually restricted to identifiable problems with constrained parameter confidence sets, here we generate prediction intervals for all three normalized loglikelihood functions in Figure \ref{fig:F4}(a)--(c).

The loglikelihood in Figure \ref{fig:F4}(a) is associated with count data, so we use it to generate a prediction interval for count data. We draw $S=500$ samples of $\vec{\theta}$ from within the 95\% confidence region, where $\bar{\ell}^{\textrm{\,c}}(\vec{\theta}\mid\mathbf{N}(t))\ge \bar{\ell}^{*}$, using a simple rejection algorithm~\cite{Simpson2026}. Candidate parameters are sampled uniformly from a user-defined region containing $\hat{\vec{\theta}}$ and retained if they satisfy the likelihood threshold. For each accepted parameter set, we solve Equations \eqref{eq:1DN}--\eqref{eq:1DP} to obtain $H N_s(x,t)$, for $s=1,\ldots,S$, where each curve gives the expected number of agents per column at location $x$ and time $t$.

\begin{figure}[htp]
  \centering
\includegraphics[width=1.0\textwidth]{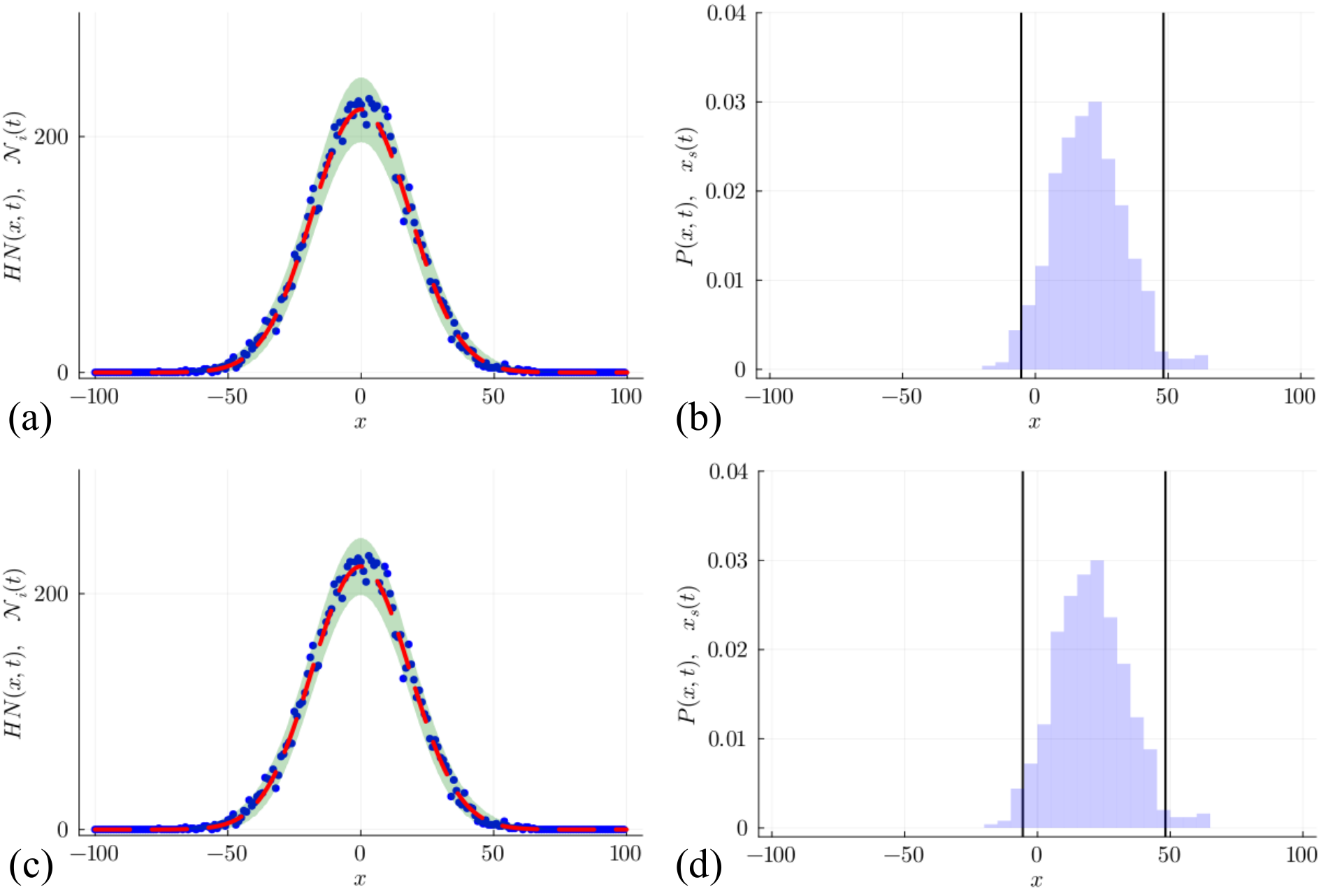}
  \caption{Likelihood-based prediction for unbiased motion using data from Figure \ref{fig:F2}(e) and Figure \ref{fig:F2}(g). (a) Count data (blue dots), MLE solution (red) and prediction interval (green) obtained from $S=500$ samples from the parameter confidence set in Figure \ref{fig:F4}(a). (b) Histogram of tagged-agent locations (blue) and prediction interval $[-5.35,48.25]$ (black vertical lines) obtained from $S=500$ samples from the confidence set in Figure \ref{fig:F4}(b). (c)--(d) Corresponding prediction intervals using the confidence set in Figure \ref{fig:F4}(c). In (d), the prediction interval is $[-5.53,48.21]$. \label{fig:F7}}
\end{figure}

To incorporate observation variability we use the binomial noise model. At each location $x$, we compute an interval $[H N_s^-(x,t),H N_s^+(x,t)]$, where the lower and upper bounds are the 0.025 and 0.975 quantiles of the binomial distribution with mean $H N_s(x,t)$. Taking the union of these intervals over $s=1,\ldots,S$ at each mesh point $x_i$ on a uniform grid over $-100\le x\le100$ with 401 mesh points gives the green prediction interval in Figure \ref{fig:F7}(a).

Figure \ref{fig:F7}(a) superimposes the noisy data and MLE solution on the prediction interval. The MLE captures the mean trend but not the stochastic variability, whereas the prediction interval encloses the data reasonably well. Specifically, it contains 397 out of 401 data points, or 99\% of the data in this realization. However, this interval is constructed from the likelihood region in Figure \ref{fig:F4}(a), where the bounds of the 95\% parameter confidence set are not determined by likelihood curvature alone. Since $K$ is not identifiable from count data, many values of $K$ match the data nearly equally well. Our accepted samples therefore satisfy both $\bar{\ell}^{\textrm{\,c}}(\vec{\theta}\mid\mathbf{N}(t))\ge \bar{\ell}^{*}$ and the imposed bound $5\le K\le15$, which reflects the truncated parameter region considered. This is a cautionary example: the prediction interval appears reasonable, but it is not based on a well-defined confidence set, so it should be interpreted carefully~\cite{Simpson2024}. A more typical use of likelihood-based prediction would focus on identifiable problems, such as the trajectory-data likelihood in Figure \ref{fig:F4}(b).

For Figure \ref{fig:F7}(a), we use rejection sampling from the approximate 95\% confidence set, although other sampling strategies are possible. For example, one could sample parameters uniformly and accept each proposal with probability $\exp[\bar{\ell}^{\textrm{\,c}}(\vec{\theta}^*\mid\mathbf{N}(t))]$ before constructing the prediction interval. Additional software available on GitHub implements this approach and gives very similar prediction intervals for the examples considered here. Other alternatives include using Laplace's approximation to sample from an associated multivariate Gaussian distribution~\cite{Simpson2025}, sampling parameter values from a uniform discretization within the 95\% confidence set, or sampling along the threshold contour where $\bar{\ell}^{\textrm{\,c}}(\vec{\theta}\mid\mathbf{N}(t))=\bar{\ell}^{*}$. For the present problems, these alternatives lead to minimal practical differences.

To make likelihood-based predictions for trajectory data, we again draw $S=500$ samples of $\vec{\theta}$ using rejection sampling. For each sample, we solve Equations \eqref{eq:1DN}--\eqref{eq:1DP} to obtain probability density functions $P_s(x,t)$, for $s=1,\ldots,S$. For the $s$th PDF, we construct the associated CDF,
\[
\phi_s(x) = \int_{-L}^{x} P_s(\xi,t)\,\textrm{d}\xi,
\]
with $\phi_s(-L)=0$. We then compute the average CDF,
\[
\Phi(x)=\dfrac{1}{S}\sum_{s=1}^{S}\phi_s(x),
\]
and define the prediction interval $[x^-,x^+]$ by $\Phi(x^-)=0.025$ and $\Phi(x^+)=0.975$. By construction, this interval has a 95\% probability of containing the location of an individual randomly sampled from the model. Figure \ref{fig:F7}(b) shows the resulting interval superimposed on the histogram of tagged-agent locations at $t=600$. The interval $[-5.35,48.25]$ contains 476 out of 500 tagged agents, or 95.20\%, in this realization.

Finally, we consider the combined-data loglikelihood in Figure \ref{fig:F4}(c). The corresponding 95\% parameter confidence set is tightly constrained around the MLE, indicating that the combined data identify $\vec{\theta}=(D,K)^\top$ reasonably precisely. We draw $S=500$ parameter samples satisfying
\[
\bar{\ell}^{\textrm{\,c+t}}(\vec{\theta}\mid\mathbf{N}(t),\mathbf{x}(t))\ge \bar{\ell}^{*}.
\]
Since this likelihood combines count and trajectory data, we use the same procedures as in Figure \ref{fig:F7}(a)--(b) to construct prediction intervals for both data types. The resulting intervals are shown in Figure \ref{fig:F7}(c)--(d). At this scale, the count-data prediction interval is visually indistinguishable from that in Figure \ref{fig:F7}(a), which was constructed from count data alone. The trajectory prediction interval in Figure \ref{fig:F7}(d) is slightly different from Figure \ref{fig:F7}(b): the trajectory-only interval is $[-5.35,48.25]$, while the combined-data interval is $[-5.53,48.21]$. The latter is slightly narrower and still contains 95.20\% of tagged-agent locations in this realization.

Overall, this exercise shows how likelihood-based prediction translates parameter uncertainty into uncertainty in measurable quantities. Care is required when the underlying parameters are poorly identified, since prediction intervals may then depend on arbitrary bounds imposed on the parameter space. While previous work has generated prediction intervals for count and density data, our results in Figure \ref{fig:F7}(b) and (d) provide, to our knowledge, the first trajectory-data prediction intervals constructed in this way.

\section{Conclusions and Future Work}

In this work we use stochastic simulations, mean-field PDE approximations and likelihood-based tools to explore the value of individual trajectory data for parameter inference and identifiability analysis. The work is motivated by the observation that experimental cell biologists and field ecologists place high value on tagging individual cells or animals within larger populations and measuring their trajectories. Rather than working with noisy experimental data, we focus on an intuitive lattice-based random walk model of population dispersal. The simulation model has three parameters: the carrying capacity of lattice sites, $\kappa>0$; the motility probability per time step for isolated agents, $M>0$; and the drift parameter, $\rho_x\in[-1,1]$. To provide a rigorous mathematical description of the simulation data, we derive a PDE-based continuum description involving three related parameters, $\vec{\theta}=(D,v,K)^\top$.

The continuum PDEs for the expected number of agents per site, $N$, and the probability density function of tagged-agent location, $P$, reveal several important features. For example, the PDE for $N$ is independent of $K$ when motility is unbiased, $\rho_x=0$. Thus, $K$ is structurally non-identifiable from count data, whereas the PDE for $P$ involves both $K$ and $D$, meaning that trajectory data can resolve this non-identifiability~\cite{Hines2014}. This provides a clear illustration of the value of trajectory data: standard count data are insensitive to $K$, whereas tagged-agent motion is sensitive to $K$. For biased motility, where $\rho_x\ne0$, the PDEs for both $N$ and $P$ depend on all three parameters, $\vec{\theta}=(D,v,K)^\top$. This indicates that, with sufficient data, all three parameters can be estimated using either count data or trajectory data.

Practical identifiability cannot be determined from the structure of the mean-field PDEs alone. We therefore use simulation data to study how parameter estimation is affected by collecting count data, trajectory data, or both. For count data, we develop a binomial loglikelihood. This differs from previous work in which $K$ was pre-estimated and the binomial coefficient did not need to be evaluated because it acted as a normalization constant~\cite{,Liu2024,Simpson2024b,Simpson2025}. Here, $K$ is unknown, so the binomial coefficient must be retained. For trajectory data, we use the new PDE model for $P$ in an inference and identifiability analysis context. For biased motility, count data lead to accurate parameter estimates and reasonably narrow confidence sets. For unbiased motility, count data estimate $D$ well, but $K$ is poorly constrained by the curvature of the loglikelihood. We also show how to combine count and trajectory data in a joint loglikelihood, leading to accurate point estimates and well-constrained confidence sets.

A major motivation for this work is that field ecologists and experimental cell biologists often collect trajectory data alongside, or instead of, count data. Since trajectory data can be costly and time-consuming to obtain, there is value in developing mathematical models and objective workflows to assess how tagging protocols affect parameter identifiability. Through simple computational experiments, we compare inferential accuracy and precision by tagging different subsets of agents in the same simulation and examining the size and shape of the resulting confidence sets. For example, tagging individuals at the leading edge gives substantially more precise estimates than tagging individuals well behind the edge of the population. This link between experimental design and parameter identifiability is important: experiments that produce poorly identifiable parameters may not justify the required time and cost. Tools that combine stochastic simulations, surrogate PDE models, likelihood-based estimation and identifiability analysis can therefore provide useful guidance for experimental design.

There are several ways to extend this work. Here we focus on an initial condition and boundary conditions in the two-dimensional random walk model that reduce the surrogate PDEs to one-dimensional equations. This simplification is relevant to many experimental studies~\cite{Cai2007,Deroulers2009,Druckenbrod2007}, but the same tools can be applied to initial conditions where the macroscopic quantities $N$ and $P$ depend on both horizontal and vertical position~\cite{Simpson2009}. In that setting, the same likelihood-based estimation, identifiability analysis and prediction workflow can be used with count data collected on a coarse mesh, together with the full two-dimensional PDEs, Equations \eqref{eq:N_PDE} and \eqref{eq:P_PDE}~\cite{Simpson2009}. The inference-identifiability-prediction workflow is otherwise unchanged.

More substantial generalizations could involve alternative coarse-graining methods that avoid the mean-field assumption~\cite{Baker2010,Ellner2001}, or lattice-free simulation frameworks~\cite{Plank2004,Plank2025a}. Here we deliberately use a lattice-based simulation model and standard mean-field PDE descriptions to keep the exposition clear. The continuum-discrete comparisons in Figure \ref{fig:F3} show that the mean-field PDEs provide accurate approximations of the random walk model in the examples considered. Our open-access GitHub software can be used to explore this comparison across other parameter values, and additional computational experiments indicate that the PDE approximation remains accurate across a broad range of conditions.

Another extension is to refine how trajectory data are used. Here we take a simple approach: for each tagged agent, we record its initial and final positions and use the PDE description of $P$ to evaluate the corresponding loglikelihood. This could be extended by recording intermediate locations along each trajectory. For example, a single trajectory could be divided into two equal time intervals, with the PDE for $P$ used to evaluate a loglikelihood contribution for each interval. This refinement is conceptually straightforward, but it requires multiple PDE solves per trajectory, whereas the present approach does not.approach is conceptually straightforward, it requires multiple PDE solves per trajectory whereas the current approach does not.

\newpage 
\section*{Appendix: Numerical Methods} \label{sec:appendix}

We solve Equations \eqref{eq:1DN}--\eqref{eq:1DP} on a uniform grid over $-L \le x \le L$ with spacing $\delta>0$, nodes $x_i=-L+(i-1)\delta$, $i=1,\ldots,\mathcal{I}$, and $\mathcal{I}=\lfloor 2L/\delta+1/2\rfloor+1$. The numerical approximations to Equations \eqref{eq:N_PDE}--\eqref{eq:P_PDE} are denoted by $N(x_i,t)$ and $P(x_i,t)$. Central differences at interior nodes give
\begin{align}
 \dfrac{\textrm{d} N(x_i,t)}{\textrm{d} t} &= -\dfrac{1}{\delta}\left[\mathcal{J}_x^{i+1/2} -  \mathcal{J}_x^{i-1/2}\right], \label{eq:DiscretisedODEb} \quad \textrm{where}, \\
\mathcal{J}_x^{i+1/2} & = -\dfrac{D}{\delta} \left(N(x_{i+1},t) - N(x_{i},t) \right) \notag \\
                      &+ \dfrac{v_x}{2} \left[N(x_{i},t)\left(1 - \dfrac{N(x_{i},t)}{K}\right)+N(x_{i+1},t)\left(1 - \dfrac{N(x_{i+1},t)}{K}\right) \right],  \\
\mathcal{J}_x^{i-1/2} & = -\dfrac{D}{\delta} \left(N(x_{i},t) - N(x_{i-1},t) \right) \notag \\
                  & + \dfrac{v_x}{2} \left[N(x_{i-1},t)\left(1 - \dfrac{N(x_{i-1},t)}{K}\right)+N(x_{i},t)\left(1 - \dfrac{N(x_{i},t)}{K}\right) \right],  
 \end{align}
for $i=2,\ldots,\mathcal{I}-1$. Forward and backward differences at $i=1$ and $i=\mathcal{I}$, respectively, with zero boundary flux, give
\begin{align}
 \dfrac{\textrm{d} N(x_1,t)}{\textrm{d} t} &=  \dfrac{D}{\delta^2}\left(N(x_2,t)-N(x_1,t)\right) - \dfrac{v_x}{\delta}N(x_2,t) \left[1 - \dfrac{N(x_2,t)}{K} \right], \quad \textrm{and}  \\
  \dfrac{\textrm{d} N(x_\mathcal{I},t)}{\textrm{d} t} &= \dfrac{D}{\delta^2}\left(N(x_{\mathcal{I}-1},t)-N(x_{\mathcal{I}},t)\right) + \dfrac{v_x}{\delta}N(x_{\mathcal{I}-1},t) \left[1 - \dfrac{N(x_{\mathcal{I}-1},t)}{K} \right].
 \end{align}

Similarly, for $P(x,t)$,
\begin{align}
 \dfrac{\textrm{d} P(x_i,t)}{\textrm{d} t} &= -\dfrac{1}{\delta}\left[\mathcal{J}_x^{i+1/2} -  \mathcal{J}_x^{i-1/2}\right], \quad \textrm{where}\\
\mathcal{J}_x^{i+1/2} =& -\dfrac{D}{2\delta} \left[\left(1 - \dfrac{N(x_{i},t)}{K} \right)+ \left(1 - \dfrac{N(x_{i+1},t)}{K} \right) \right] \left(P(x_{i+1},t)-P(x_{i},t) \right) \notag \\
                &-\dfrac{D}{2\delta K}\left(P(x_{i},t) + P(x_{i+1},t) \right)\left(N(x_{i+1},t)- N(x_{i},t) \right) \notag  \\
&+\dfrac{v_x}{2} \left[P(x_{i},t)\left(1 - \dfrac{N(x_{i},t)}{K} \right)+P(x_{i+1},t)\left(1 - \dfrac{N(x_{i+1},t)}{K} \right) \right],\\
\mathcal{J}_x^{i-1/2} =& -\dfrac{D}{2\delta} \left[\left(1 - \dfrac{N(x_{i},t)}{K} \right)+ \left(1 - \dfrac{N(x_{i-1},t)}{K} \right) \right] \left(P(x_{i},t)-P(x_{i-1},t) \right)  \notag \\
&-\dfrac{D}{2\delta K}\left(P(x_{i},t) + P(x_{i-1},t) \right)\left(N(x_{i},t) - N(x_{i-1},t) \right)  \notag \\
&+\dfrac{v_x}{2} \left[P(x_{i},t)\left(1 - \dfrac{N(x_{i},t)}{K} \right)+P(x_{i-1},t)\left(1 - \dfrac{N(x_{i-1},t)}{K} \right) \right],
 \end{align}
for $i=2,\ldots,\mathcal{I}-1$. The zero-flux boundary equations are
\begin{align}
& \dfrac{\textrm{d} P(x_1,t)}{\textrm{d} t}  =  \dfrac{D}{\delta^2}\left[1 - \dfrac{N(x_2,t)}{K} \right]\left(P(x_2,t)-P(x_1,t)\right) \notag  \\
                                            &+\dfrac{D}{\delta^2 K}P(x_2,t)\left(N(x_2,t) -N(x_1,t)\right) 
                                            - \dfrac{v_x}{\delta}P(x_2,t) \left[1 - \dfrac{N(x_2,t)}{K} \right],  \quad \textrm{and} \\
  &\dfrac{\textrm{d} P(x_\mathcal{I},t)}{\textrm{d} t} = \dfrac{D}{\delta^2}\left[1 - \dfrac{N(x_{\mathcal{I}-1},t)}{K} \right]\left(P(x_{\mathcal{I}-1},t)-P(x_{\mathcal{I}},t)\right) \notag   \\
                                            &+\dfrac{D}{\delta^2 K}P(x_{\mathcal{I}-1},t)\left(N(x_{\mathcal{I}-1},t) -N(x_{\mathcal{I}},t)\right) - \dfrac{v_x}{\delta}P(x_{\mathcal{I}-1},t) \left[1 - \dfrac{N(x_{\mathcal{I}-1},t)}{K} \right]\label{eq:DiscretisedODEe}.
 \end{align}
 
The resulting $2\mathcal{I}$ coupled ordinary differential equations, Equations \eqref{eq:DiscretisedODEb}--\eqref{eq:DiscretisedODEe}, are integrated in Julia using DifferentialEquations.jl~\cite{Rackauckas2017}. We use Heun's method with standard tolerances and automatic time stepping. All results use $\delta=0.5$, which gives grid-independent results for the problems considered.

\newpage

\noindent
\textbf{Data Accessibility} Julia implementations within Jupyter notebooks for all computations are available on GitHub at \href{https://github.com/ProfMJSimpson/RandomWalkTrajectories}{https://github.com/ProfMJSimpson/RandomWalkTrajectories}.

\noindent
\textbf{Funding} This work is partly supported by the Australian Research Council (DP230100025) and the Marsden Fund (24-UOC-020).

\bibliography{references}

@book{AbramowitzS1964,
  author    = {Milton Abramowitz and Irene A. Stegun},
  title     = {Handbook of Mathematical Functions with Formulas, Graphs, and Mathematical Tables},
  series    = {National Bureau of Standards Applied Mathematics Series},
  volume    = {55},
  year      = {1964}
}

@article{Anguige2009,
  title   = {A one-dimensional model of cell diffusion and aggregation, incorporating volume filling and cell-to-cell adhesion},
  author  = {Anguige, K. and Schmeiser, C.},
  journal = {Journal of Mathematical Biology},
  volume  = {58},
  number  = {3},
  pages   = {395--427},
  year    = {2009},
  doi     = {10.1007/s00285-008-0197-8}
}

@article{Baker2010,
  author  = {Baker, Ruth E. and Simpson, Matthew J.},
  title   = {Correcting mean-field approximations for birth-death-movement processes},
  journal = {Physical Review E},
  year    = {2010},
  volume  = {82},
  number  = {4},
  pages   = {041905},
  doi     = {10.1103/PhysRevE.82.041905}
}

@article{Baker2022,
  author  = {Baker, Phillip A. and Ibarra-Garc{\'i}a-Padilla, Rodrigo and Venkatesh, Akshaya and Singleton, Eileen W. and Uribe, Rosa A.},
  title   = {In toto imaging of early enteric nervous system development reveals that gut colonization is tied to proliferation downstream of {Ret}},
  journal = {Development},
  year    = {2022},
  volume  = {149},
  number  = {21},
  pages   = {dev200668},
  doi     = {10.1242/dev.200668}
}

@book{Berg1983,
  author    = {Berg, Howard C.},
  title     = {Random Walks in Biology},
  edition   = {Expanded},
  publisher = {Princeton University Press},
  year      = {1983}
}

@article{Browning2017,
  title   = {A {B}ayesian computational approach to explore the optimal duration of a cell proliferation assay},
  author  = {Browning, Alexander P. and McCue, Scott W. and Simpson, Matthew J.},
  journal = {Bulletin of Mathematical Biology},
  volume  = {79},
  pages   = {1888--1906},
  year    = {2017},
  doi     = {10.1007/s11538-017-0311-4}
}

@article{Bruna2012,
  title   = {Excluded-volume effects in the diffusion of hard spheres},
  author  = {Bruna, M. and Chapman, S. J.},
  journal = {Physical Review E},
  volume  = {85},
  number  = {1},
  pages   = {011103},
  year    = {2012},
  doi     = {10.1103/PhysRevE.85.011103}
}

@article{Cai2007,
  title   = {Multi-scale modeling of a wound-healing cell migration assay},
  author  = {Cai, Anna Q. and Landman, Kerry A. and Hughes, Barry D.},
  journal = {Journal of Theoretical Biology},
  volume  = {245},
  number  = {3},
  pages   = {576--594},
  year    = {2007},
  doi     = {10.1016/j.jtbi.2006.10.024}
}

@article{Callaghan2006,
  title   = {A stochastic model for wound healing},
  author  = {Callaghan, Thomas and Khain, Evgeniy and Sander, Leonard M. and Ziff, Robert M.},
  journal = {Journal of Statistical Physics},
  volume  = {122},
  number  = {5},
  pages   = {909--924},
  year    = {2006},
  doi     = {10.1007/s10955-006-9022-1}
}

@article{Cavagna2013,
  author  = {Cavagna, Andrea and Duarte Queir{\'o}s, Silvio M. and Giardina, Irene and Stefanini, Fabio and Viale, Massimiliano},
  title   = {Diffusion of individual birds in starling flocks},
  journal = {Proceedings of the Royal Society B: Biological Sciences},
  year    = {2013},
  volume  = {280},
  number  = {1756},
  pages   = {20122484},
  doi     = {10.1098/rspb.2012.2484}
}

@article{Chis2011,
  author  = {Chi{\c{s}}, Oana and Banga, Julio R. and Balsa-Canto, Eva},
  title   = {Structural identifiability of systems biology models: a critical comparison of methods},
  journal = {PLoS ONE},
  year    = {2011},
  volume  = {6},
  number  = {11},
  pages   = {e27755},
  doi     = {10.1371/journal.pone.0027755}
}

@article{Chowdhury2005,
  title   = {Physics of transport and traffic phenomena in biology: From molecular motors and cells to organisms},
  author  = {Chowdhury, D. and Schadschneider, A. and Nishinari, K.},
  journal = {Physics of Life Reviews},
  volume  = {2},
  pages   = {318--352},
  year    = {2005},
  doi     = {10.1016/j.plrev.2005.09.001}
}

@article{Codling2008,
  title   = {Random walk models in biology},
  author  = {Codling, E. A. and Plank, M. J. and Benhamou, S.},
  journal = {Journal of the Royal Society Interface},
  volume  = {5},
  pages   = {813--834},
  year    = {2008},
  doi     = {10.1098/rsif.2008.0014}
}

@article{Crossley2023,
  author  = {Crossley, Rebecca M. and Maini, Philip K. and Lorenzi, Tommaso and Baker, Ruth E.},
  title   = {Traveling waves in a coarse-grained model of volume-filling cell invasion: Simulations and comparisons},
  journal = {Studies in Applied Mathematics},
  volume  = {151},
  number  = {4},
  pages   = {1471--1497},
  year    = {2023},
  doi     = {10.1111/sapm.12635}
}

@article{Crossley2024,
  author  = {Crossley, Rebecca M. and Painter, Kevin J. and Lorenzi, Tommaso and Maini, Philip K. and Baker, Ruth E.},
  title   = {Phenotypic switching mechanisms determine the structure of cell migration into extracellular matrix under the `go-or-grow' hypothesis},
  journal = {Mathematical Biosciences},
  volume  = {374},
  pages   = {109240},
  year    = {2024},
  doi     = {10.1016/j.mbs.2024.109240}
}

@article{Deroulers2009,
  title   = {Modeling tumor cell migration: From microscopic to macroscopic models},
  author  = {Deroulers, C. and Aubert, M. and Badoual, M. and Grammaticos, B.},
  journal = {Physical Review E},
  volume  = {79},
  number  = {3},
  pages   = {031917},
  year    = {2009},
  doi     = {10.1103/PhysRevE.79.031917}
}

@article{DiazSeoane2022,
  author  = {D{\'{\i}}az-Seoane, Sergio and Rey Barreiro, Xabier and Villaverde, Alejandro F.},
  title   = {{STRIKE-GOLDD} 4.0: user-friendly, efficient analysis of structural identifiability and observability},
  journal = {Bioinformatics},
  year    = {2022},
  volume  = {39},
  number  = {1},
  pages   = {btac748},
  doi     = {10.1093/bioinformatics/btac748}
}

@article{Druckenbrod2005,
  author  = {Druckenbrod, Noah R. and Epstein, Miles L.},
  title   = {The pattern of neural crest advance in the cecum and colon},
  journal = {Developmental Biology},
  volume  = {287},
  number  = {1},
  pages   = {125--133},
  year    = {2005},
  doi     = {10.1016/j.ydbio.2005.08.040}
}

@article{Druckenbrod2007,
  author  = {Druckenbrod, Noah R. and Epstein, Miles L.},
  title   = {Behavior of enteric neural crest-derived cells varies with respect to the migratory wavefront},
  journal = {Developmental Dynamics},
  volume  = {236},
  number  = {1},
  pages   = {84--92},
  year    = {2007},
  doi     = {10.1002/dvdy.20974}
}

@book{EdelsteinKeshet2005,
  author    = {Leah Edelstein-Keshet},
  title     = {Mathematical Models in Biology},
  volume    = {46},
  year      = {2005},
  publisher = {Society for Industrial and Applied Mathematics},
  address   = {Philadelphia, PA},
  isbn      = {978-0-89871-554-5}
}

@article{Ellner2001,
  title   = {Pair approximation for lattice models with multiple interaction scales},
  author  = {Ellner, Stephen P.},
  journal = {Journal of Theoretical Biology},
  volume  = {210},
  number  = {4},
  pages   = {435--447},
  year    = {2001},
  doi     = {10.1006/jtbi.2001.2322}
}

@inproceedings{Froehlich2014,
  author    = {Fr{\"o}hlich, Fabian and Theis, Fabian J. and Hasenauer, Jan},
  title     = {Uncertainty analysis for non-identifiable dynamical systems: profile likelihoods, bootstrapping and more},
  booktitle = {Computational Methods in Systems Biology},
  year      = {2014},
  pages     = {61--72},
  publisher = {Springer},
  doi = {10.1007/978-3-319-12982-2\_5}
}

@article{Harada1994,
  title   = {Lattice population dynamics for plants with dispersing seeds and vegetative propagation},
  author  = {Harada, Yuko and Iwasa, Yoh},
  journal = {Researches on Population Ecology},
  volume  = {36},
  pages   = {237--249},
  year    = {1994},
  doi     = {10.1007/BF02514940}
}

@Article{Harris2013,
  author  = {Harris, Keith J. and Blackwell, Paul G.},
  title   = {Flexible continuous-time modelling for heterogeneous animal movement},
  journal = {Ecological Modelling},
  year    = {2013},
  volume  = {255},
  pages   = {29--37},
  doi     = {10.1016/j.ecolmodel.2013.01.020}
}

@article{Hines2014,
  title   = {Determination of parameter identifiability in nonlinear biophysical models: A Bayesian approach},
  author  = {Hines, Keegan E. and Middendorf, Thomas R. and Aldrich, Richard W.},
  journal = {Journal of General Physiology},
  volume  = {143},
  number  = {4},
  pages   = {401--416},
  year    = {2014},
  doi     = {10.1085/jgp.201311116}
}

@book{Kot2001,
  author    = {Mark Kot},
  title     = {Elements of Mathematical Ecology},
  year      = {2001},
  publisher = {Cambridge University Press},
  address   = {Cambridge},
  isbn      = {9780521001502}
}

@article{Lambert2018,
  title   = {Bayesian inference of agent-based models: a tool for studying kidney branching morphogenesis},
  author  = {Lambert, B. and MacLean, A. L. and Fletcher, A. G. and Coombes, A. N. and Little, M. H. and Byrne, H. M.},
  journal = {Journal of Mathematical Biology},
  year    = {2018},
  volume  = {76},
  pages   = {1673--1697},
  doi     = {10.1007/s00285-018-1208-z}
}

@article{Ligon2018,
  author  = {Ligon, Thomas S. and Fr{\"o}lich, Fabian and Chi{\c{s}}, Oana and Banga, Julio R. and Balsa-Canto, Eva and Hasenauer, Jan},
  title   = {{GenSSI} 2.0: multi-experimental structural identifiability analysis of {SBML} models},
  journal = {Bioinformatics},
  year    = {2018},
  volume  = {34},
  number  = {8},
  pages   = {1421--1423},
  doi     = {10.1093/bioinformatics/btx735}
}

@article{Liu2024,
  author  = {Liu, Yihan and Warne, David J. and Simpson, Matthew J.},
  title   = {Likelihood-based inference, identifiability, and prediction using count data from lattice-based random walk models},
  journal = {Physical Review E},
  year    = {2024},
  volume  = {110},
  number  = {4},
  pages   = {044405},
  doi     = {10.1103/PhysRevE.110.044405}
}

@article{Miles2025,
  author  = {Miles, Christopher E.},
  title   = {Incorporating spatial diffusion into models of bursty stochastic transcription},
  journal = {Journal of the Royal Society Interface},
  year    = {2025},
  volume  = {22},
  number  = {225},
  pages   = {20240739},
  doi     = {10.1098/rsif.2024.0739}
}

@article{Morales2022,
  author  = {Morales, Juan Manuel and Mor{\'a}n L{\'o}pez, Teresa},
  title   = {Mechanistic models of seed dispersal by animals},
  journal = {Oikos},
  year    = {2022},
  volume  = {2022},
  number  = {2},
  pages   = {e08328},
  doi     = {10.1111/oik.08328}
}

@article{Mort2016,
  title   = {Reconciling diverse mammalian pigmentation patterns with a fundamental mathematical model},
  author  = {Mort, Richard L. and Ross, Robert J. H. and Hainey, Kirsten J. and Harrison, Olivia J. and Keighren, Margaret A. and Landini, Gabriel and Baker, Ruth E. and Painter, Kevin J. and Jackson, Ian J. and Yates, Christian A.},
  journal = {Nature Communications},
  volume  = {7},
  pages   = {10288},
  year    = {2016},
  doi     = {10.1038/ncomms10288}
}

@book{Murray2002,
  author    = {Murray, J. D.},
  title     = {Mathematical Biology I: An Introduction},
  edition   = {3},
  publisher = {Springer},
  address   = {New York},
  year      = {2002},
  series    = {Interdisciplinary Applied Mathematics},
  volume    = {17},
  doi       = {10.1007/b98868}
}

@article{Ovaskainen2008,
  author  = {Ovaskainen, Otso and Smith, A. D. and Osborne, J. L. and Reynolds, D. R. and Carreck, N. L. and Martin, A. P. and Niitep{\~o}ld, Kristjan and Hanski, Ilkka},
  title   = {Tracking butterfly movements with harmonic radar reveals an effect of population age on movement distance},
  journal = {Proceedings of the National Academy of Sciences of the United States of America},
  year    = {2008},
  volume  = {105},
  number  = {49},
  pages   = {19090--19095},
  doi     = {10.1073/pnas.0802066105}
}

@book{Okubo2001,
  author    = {Okubo, Akira and Levin, Simon A.},
  title     = {Diffusion and Ecological Problems: Modern Perspectives},
  edition   = {2},
  publisher = {Springer},
  address   = {New York},
  year      = {2001},
  doi       = {10.1007/978-1-4757-4978-6}
}

@article{Painter2002,
  author  = {Painter, Kevin J. and Hillen, Thomas},
  title   = {Volume-filling and quorum-sensing in models for chemosensitive movement},
  journal = {Canadian Applied Mathematics Quarterly},
  volume  = {10},
  number  = {4},
  pages   = {501--543},
  year    = {2002}
}

@article{Patterson2008,
  author  = {Patterson, Toby A. and Thomas, Len and Wilcox, Chris and Ovaskainen, Otso and Matthiopoulos, Jason},
  title   = {State-space models of individual animal movement},
  journal = {Trends in Ecology \& Evolution},
  year    = {2008},
  volume  = {23},
  pages   = {87--94},
  doi     = {10.1016/j.tree.2007.10.009}
}

@book{Pawitan2001,
  title  = {In All Likelihood: Statistical Modelling and Inference Using Likelihood},
  author = {Pawitan, Yudi},
  year   = {2001},
  doi    = {10.1093/oso/9780198507659.001.0001},
  series = {Oxford Science Publications}
}

@article{Perry2015,
  author  = {Perry, Clint J. and Søvik, Eirik and Myerscough, Mary R. and Barron, Andrew B.},
  title   = {Rapid behavioral maturation accelerates failure of stressed honey bee colonies},
  journal = {Proceedings of the National Academy of Sciences of the United States of America},
  volume  = {112},
  number  = {11},
  pages   = {3427--3432},
  year    = {2015},
  doi     = {10.1073/pnas.1422089112}
}

@article{Plank2004,
  title   = {Lattice and non-lattice models of tumour angiogenesis},
  author  = {Plank, M. J. and Sleeman, B. D.},
  journal = {Bulletin of Mathematical Biology},
  volume  = {66},
  number  = {6},
  pages   = {1785--1819},
  year    = {2004},
  doi     = {10.1016/j.bulm.2004.04.001}
}

@article{Plank2025a,
  title   = {Random walk models in the life sciences: including births, deaths and local interactions},
  author  = {Plank, M. J. and Simpson, M. J. and Baker, R. E.},
  journal = {Journal of the Royal Society Interface},
  volume  = {22},
  number  = {222},
  pages   = {20240422},
  year    = {2025},
  doi     = {10.1098/rsif.2024.0422}
}

@article{Plank2026,
  author  = {Plank, Michael J. and Simpson, Matthew J.},
  title   = {Continuum models describing probabilistic motion of tagged agents in exclusion processes},
  journal = {Physical Review E},
  year    = {2026},
  volume  = {113},
  pages   = {014137},
  doi     = {10.1103/lzqy-n5mw}
}

@article{Rackauckas2017,
  author  = {Rackauckas, Christopher and Nie, Qing},
  title   = {{DifferentialEquations.jl}: A performant and feature-rich ecosystem for solving differential equations in Julia},
  journal = {Journal of Open Research Software},
  volume  = {5},
  number  = {1},
  pages   = {15},
  year    = {2017},
  doi     = {10.5334/jors.151}
}

@article{Raue2009,
  author  = {Raue, Andreas and Kreutz, Clemens and Maiwald, Thomas and Bachmann, Julie and Schilling, Marcel and Klingm{\"u}ller, Ursula and Timmer, Jens},
  title   = {Structural and practical identifiability analysis of partially observed dynamical models by exploiting the profile likelihood},
  journal = {Bioinformatics},
  year    = {2009},
  volume  = {25},
  number  = {15},
  pages   = {1923--1929},
  doi     = {10.1093/bioinformatics/btp358}
}

@article{Royston2007,
  author  = {Patrick Royston},
  title   = {Profile likelihood for estimation and confidence intervals},
  journal = {The Stata Journal},
  volume  = {7},
  number  = {3},
  pages   = {376--387},
  year    = {2007},
  doi     = {10.1177/1536867X0700700305}
}

@article{Seber1986,
  author  = {Seber, G. A. F.},
  title   = {A review of estimating animal abundance},
  journal = {Biometrics},
  volume  = {42},
  number  = {2},
  pages   = {267--292},
  year    = {1986},
  doi     = {10.2307/2531049}
}

@article{Simpson2009,
  author  = {Simpson, Matthew J. and Landman, Kerry A. and Hughes, Barry D.},
  title   = {Multi-species simple exclusion processes},
  journal = {Physica A: Statistical Mechanics and its Applications},
  volume  = {388},
  number  = {4},
  pages   = {399--406},
  year    = {2009},
  doi     = {10.1016/j.physa.2008.10.038}
}

@article{Simpson2013,
  title   = {Quantifying the roles of cell motility and cell proliferation in a circular barrier assay},
  author  = {Simpson, M. J. and Treloar, K. K. and Binder, B. J. and Haridas, P. and Manton, K. J. and Leavesley, D. I. and McElwain, D. L. S. and Baker, R. E.},
  journal = {Journal of the Royal Society Interface},
  volume  = {10},
  number  = {82},
  pages   = {20130007},
  year    = {2013},
  doi     = {10.1098/rsif.2013.0007}
}

@article{Simpson2024,
  author  = {Simpson, Matthew J. and Maclaren, Oliver J.},
  title   = {Making predictions using poorly identified mathematical models},
  journal = {Bulletin of Mathematical Biology},
  year    = {2024},
  volume  = {86},
  number  = {7},
  pages   = {80},
  doi     = {10.1007/s11538-024-01294-0}
}

@article{Simpson2024b,
  author  = {Simpson, Matthew J. and Murphy, Ryan J. and Maclaren, Oliver J.},
  title   = {Modelling count data with partial differential equation models in biology},
  journal = {Journal of Theoretical Biology},
  volume  = {580},
  pages   = {111732},
  year    = {2024},
  doi     = {10.1016/j.jtbi.2024.111732}
}

@article{Simpson2025,
  title   = {Inference and prediction for stochastic models of biological populations undergoing migration and proliferation},
  author  = {Simpson, Matthew J. and Plank, Michael J.},
  journal = {Journal of the Royal Society Interface},
  year    = {2025},
  volume  = {22},
  number  = {231},
  pages   = {20250536},
  doi     = {10.1098/rsif.2025.0536}
}

@article{Simpson2026,
  title   = {Parameter identifiability, parameter estimation and model prediction for differential equation models},
  author  = {Simpson, M. J. and Baker, R. E.},
  journal = {SIAM Review},
  volume  = {68},
  pages   = {153--171},
  year    = {2026},
  doi     = {10.1137/24M1667968},
}

@article{Skellam1951,
  author  = {Skellam, J. G.},
  title   = {Random dispersal in theoretical populations},
  journal = {Biometrika},
  volume  = {38},
  number  = {1--2},
  pages   = {196--218},
  year    = {1951},
  doi     = {10.1093/biomet/38.1-2.196}
}

@article{Siekmann2012,
  author  = {Siekmann, Ivo and Sneyd, James and Crampin, Edmund J.},
  title   = {{MCMC} can detect nonidentifiable models},
  journal = {Biophysical Journal},
  year    = {2012},
  volume  = {103},
  number  = {11},
  pages   = {2275--2286},
  doi     = {10.1016/j.bpj.2012.10.024}
}

@article{Stokes1991a,
  author  = {Stokes, Cynthia L. and Lauffenburger, Douglas A. and Williams, Stuart K.},
  title   = {Migration of individual microvessel endothelial cells: stochastic model and parameter measurement},
  journal = {Journal of Cell Science},
  volume  = {99},
  number  = {2},
  pages   = {419--430},
  year    = {1991},
  doi     = {10.1242/jcs.99.2.419}
}

@Article{Taylor2015,
  author  = {Taylor, P. R. and Yates, C. A. and Simpson, M. J. and Baker, R. E.},
  title   = {Reconciling transport models across scales: The role of volume exclusion},
  journal = {Physical Review E},
  year    = {2015},
  volume  = {92},
  number  = {4},
  pages   = {040701},
  doi     = {10.1103/PhysRevE.92.040701},
}

@article{Treloar2013,
  author  = {Treloar, Katrina K. and Simpson, Matthew J. and Haridas, Parvathi and Manton, Kerry J. and Leavesley, David I. and McElwain, D. L. Sean and Baker, Ruth E.},
  title   = {Multiple types of data are required to identify the mechanisms influencing the spatial expansion of melanoma cell colonies},
  journal = {BMC Systems Biology},
  volume  = {7},
  pages   = {137},
  year    = {2013},
  doi     = {10.1186/1752-0509-7-137}
}

@article{Vardeman1992,
  author  = {Vardeman, Stephen B.},
  title   = {What about the other intervals?},
  journal = {The American Statistician},
  volume  = {46},
  number  = {3},
  pages   = {193--197},
  year    = {1992},
  doi     = {10.1080/00031305.1992.10475882}
}

@article{Wilks1938,
  title   = {The large-sample distribution of the likelihood ratio for testing composite hypotheses},
  author  = {Wilks, S. S.},
  journal = {The Annals of Mathematical Statistics},
  volume  = {9},
  number  = {1},
  pages   = {60--62},
  year    = {1938},
  doi     = {10.1214/aoms/1177732360}
}

@article{Young2001,
  author  = {Young, Heather M. and Hearn, C. J. and Farlie, P. G. and Canty, A. J. and Thomas, P. Q. and Newgreen, D. F.},
  title   = {{GDNF} is a chemoattractant for enteric neural cells},
  journal = {Developmental Biology},
  volume  = {229},
  number  = {2},
  pages   = {503--516},
  year    = {2001},
  doi     = {10.1006/dbio.2000.0100}
}
\end{document}